# A multilevel backbone extraction framework

Sanaa Hmaida[1], Hocine Cherifi[2] and Mohammed El Hassouni[3*]

*Correspondence:
mohamed.elhassouni@flsh.um5.ac.ma

[1] FLSH, FSR, LRIT, Mohammed V University in Rabat, Rabat, Morocco
[2] ICB UMR 6303 CNRS, University of Burgundy, 21000 Dijon, France
[3] FLSH, Mohammed V University in Rabat, Rabat, Morocco

## Abstract

As networks grow in size and complexity, backbones become an essential network representation. Indeed, they provide a simplified yet informative overview of the underlying organization by retaining the most significant and structurally influential connections within a network. Network heterogeneity often results in complex and intricate structures, making it challenging to identify the backbone. In response, we introduce the Multilevel Backbone Extraction Framework, a novel approach that diverges from conventional backbone methodologies. This generic approach prioritizes the mesoscopic organization of networks. First, it splits the network into homogeneous-density components. Second, it extracts independent backbones for each component using any classical Backbone technique. Finally, the various backbones are combined. This strategy effectively addresses the heterogeneity observed in network groupings. Empirical investigations on real-world networks underscore the efficacy of the Multilevel Backbone approach in preserving essential network structures and properties. Experiments demonstrate its superiority over classical methods in handling network heterogeneity and enhancing network integrity. The framework is adaptable to various types of networks and backbone extraction techniques, making it a versatile tool for network analysis and backbone extraction across diverse network applications.

## Introduction

In recent years, complex networks have emerged as an ideal tool for deciphering and understanding complex systems (Torre 2017; Costa et al. 2011; Mourchid et al. 2019; Qureshi et al. 2021; Rajeh et al. 2023). A network is a set of nodes connected by edges representing binary interactions, and can model any complex system (Heitzig 2011). Sparsity is particularly advantageous in networks as it reduces computational complexity in various analytical techniques. With the ever-increasing volume of data, network analysis has become more challenging, necessitating methods that retain relevant information while reducing network size. Backbones are crucial in this context, simplifying networks to their essential structure by filtering out non-critical edges and nodes (Shin 2013; Zhang 2018). These backbones find applications across diverse fields, such as social, transportation, biological, and telecommunication networks (Liang 2021). Researchers have developed techniques to condense networks, categorizable into structural and statistical approaches. Structural methods maintain the network's essential topological properties by removing nodes or links (Yassin et al. 2024a; Ghalmane et al. 2023), while statistical methods discard edges or nodes based on statistical insignificance (Yassin et al. 2024b). Structural approaches often require prior knowledge of the network's characteristics, but statistical methods can be applied to any network regardless of its structure. Backbone extraction, an active research field, focuses on distilling networks to their most significant components (Cao 2019; Yassin et al. 2023b; Ghalmane et al. 2020a). This process can uncover crucial patterns, communities, and other network characteristics (Philippe 2010).

The component structure is a key property of real-world networks, which refers to how nodes are organized into distinct groups or clusters based on their connectivity (Newman and Girvan 2004; Clauset et al. 2004). In particular, many real-world networks exhibit a modular structure, where nodes within a module tend to be highly interconnected with each other but only weakly connected to nodes in different modules (Ravasz et al. 2002; Newman 2006). There are various methods for analyzing the component structure of networks, such as community detection algorithms that identify clusters of highly connected nodes, or centrality measures that quantify the importance of individual nodes or edges (Fortunato 2010; Freeman 1977). By studying the component structure of real-world networks, researchers can gain a deeper understanding of the complex patterns of connectivity that underlie many natural and engineered systems, from social networks to transportation networks to biological systems (Estrada 2012).

In this work, we present a novel approach for extracting a network backbone based on its mesoscopic properties. The mesoscopic structure of a network refers to the organization of node subgroups within the network, which can be identified through the community or core-periphery structure. Our approach is built upon the recently introduced component structure, which has proven to be flexible (Diop et al. 2021, 2022a, 2023a, 2022b, 2023e, b, c, d). However, other mesoscopic representations can be used as well. To extract the backbone of a network, we propose a generic framework that can utilize any backbone extraction technique developed for weighted networks. We conduct experiments on real-world networks using two prominent techniques: the Global Threshold and the Disparity Filter. We use several measures such as global properties, distributions, distances, and mesoscopic properties to demonstrate the effectiveness of leveraging the component structure to extract the backbone. This approach can reveal important insights into complex networks' underlying organization and behavior.

The article is structured as follows: "Related work" section discusses related work, "Background" section provides the research background, and "Multilevel backbone extraction framework based on the component structure" section details our backbone extraction framework. In "Data and methods" section describes the data and methods, while in "Experimental results and discussion" section interprets the results from the Global Threshold and Disparity Filter applications. In "Conclusion" section concludes the paper.

## Related work

In network analysis, the network backbone refers to a simplified version of the original network that retains essential and relevant information while eliminating noise and irrelevant details. Extracting a backbone enables more streamlined management and comprehension of complex networks. The research on backbone extraction primarily concentrates on two types of networks: mono-mode networks and bipartite networks. This study focuses explicitly on mono-mode networks, where backbone extraction is accomplished through coarse-graining or filter-based approaches (Ghalmane et al. 2021). Coarse-graining methods aim to identify nodes with shared characteristics, aggregating them into communities or groups represented as single nodes. This process reduces the overall size of the network while preserving its essential properties. Typically, the assembled nodes exhibit dense interconnectivity within their community, distinguishing them from nodes outside the group (Ahn et al. 2010). For instance, Gfeller and De Los Rios (2007) describe in their study how they maintain random walks and group nodes with similar neighbors, making them indistinguishable in terms of random walk perspectives. On the other hand, filter-based approaches employ a bottom-up strategy to extract the network backbone (Ghalmane et al. 2021). They define specific statistical features for nodes or links and use them as criteria to determine which nodes or links should be retained or discarded from the original network. Consequently, the resulting backbone comprises only the most relevant nodes and links based on the defined properties (Boccaletti et al. 2006).

Structural and statistical methods are two prominent approaches for extracting backbones from networks in network analysis. Statistical backbones employ statistical measures to identify the network's most significant or informative elements. These methods leverage node degree, eigenvector centrality, local clustering coefficient, and edge betweenness to filter out less important nodes or edges and retain the most influential ones (Yassin et al. 2022b). Statistical backbones provide insights into the importance or relevance of individual network components based on their statistical characteristics. These methods can be categorized based on the type of information utilized for the filtering process, which can be global, local, or a combination of both. A previous study by Serrano et al. (2009) used local measures to extract the backbone. They introduce a filtering method that extracts the important connections in complex multiscale networks. The approach considers small-scale interactions, maintaining edges that signify statistically relevant deviations from a null model. It is evaluated against alternative techniques and deployed on real-world networks. The results revealed the efficacy of the Disparity Filter algorithm in extracting dense subnetworks from weighted networks with missing links. However, the algorithm was limited because it only applied to undirected graphs and graphs with no cycles. Additionally, it was assumed that the weights of the edges were uniformly distributed across the different subsets of the network. Other filtering methods employ global measures, such as the "link salience" approach (Grady et al. 2012). This technique involved constructing the shortest path tree, which summarizes the shortest connections from a reference node to the rest of the network. An average shortest-path tree matrix $S$ is calculated, where each element $s_{ij}$ represents the number of shortest-path trees in which the link $(i, j)$ is included. Only edges with values above a certain threshold are retained, forming the network's backbone. The robust approach worked well on many empirical networks, enabling the prediction of essential features of contagion phenomena and offering a better understanding of the hidden universal features in complex networks. Other types of filtering methods combine both local and global measures. For example, the h-Backbone technique, which entails three phases, was suggested by Zhang et al. (2018). The first step was determining each link's bridge measure, calculated by dividing its betweenness by the total number of nodes in the network. The h-bridge ($h_b$) was then determined as the largest number of links with a bridge measure greater than or equal to $h_b$. Second, the h-strength ($h_s$) was identified as the largest number of links with strength greater than or equal to $h_s$. Finally, the links that satisfy the h-bridge and h-strength criteria were merged to form the h-backbone, which included significant links connecting the network and high-strength links located in the network's core.

On the other hand, structural backbones focus on identifying the core structure of a network by capturing its essential connections and relationships. These methods simplify the network representation while preserving its overall structural properties. Techniques such as k-core decomposition, modularity-based methods, and community detection algorithms are commonly used to extract the structural backbone (Dai et al. 2018). Recent advancements in community detection algorithms have significantly contributed to understanding complex networks. Asmi et al. (2022) introduced the greedy coupled-seeds expansion method for overlapping community detection, which efficiently identifies overlapping communities by expanding seed nodes based on their connectivity. This method has demonstrated superior performance in large-scale social networks due to its balance of computational efficiency and detection accuracy. In another study, Asmi et al. (2017) developed a large-scale community detection algorithm based on a new dissimilarity measure. This algorithm addresses scalability issues in large networks by introducing a novel way to quantify dissimilarity between nodes, facilitating accurate community boundary identification. Both methods offer robust solutions for detecting community structures in various types of networks, enhancing the overall understanding of network dynamics. Ghalmane et al. recently proposed node-filtering techniques based on the network's community structure (Ghalmane et al. 2020a). They presented two algorithms: one preserves overlapping nodes and network hubs, while the other retains overlapping nodes and their one-step neighbors to form the backbone. These algorithms outperform the widely used disparity filter, emphasizing the importance of community structure in preserving essential network information while reducing its size. The experiments were performed on real-world weighted networks from various domains and compared with the disparity filter. Results showed that the proposed methods were almost identical and more effective than the disparity filter in preserving the relevant nodes and connections. The effectiveness of the three backbones is also compared in terms of information gateway, connectedness, and link strength, and results showed that the proposed methods were effective in uncovering the most relevant components of the network. Building on these insights, Rajeh et al. (2022) proposed the "modularity vitality backbone" algorithm, leveraging the network's community structure. This technique evaluated the node contribution to modularity, filtering out nodes with the lowest contributions to the quality measure of the community structure. The remaining nodes constituted the backbone. Experimental results demonstrated that the modularity vitality backbone performed well in terms of weighted modularity, average weighted degree, and average link weight compared to alternative methods. However, it did not specifically retain nodes contributing to information spreading efficiency. Instead, it strategically preserved nodes and their edges that enhance the network's modularity.

In light of these studies, we propose a Multilevel Backbone Extraction Framework that utilizes any technique designed for extracting backbones in weighted networks and exploits the network's component structure to preserve mesoscopic features.

## Background

### Mesoscopic representation

The mesoscopic representation of a network provides an intermediate-scale analysis that bridges the macroscopic level of the entire network and the microscopic level of individual nodes (Barrat et al. 2008). It focuses on subgroups of nodes such as communities or clusters, which allows for a detailed study of the network's structure and dynamics.

### Community structure

Community structure refers to dividing a complex network into distinct groups or communities, where nodes within each community are more densely connected than nodes outside the community (Cherifi et al. 2019). In simpler terms, it's the organization of nodes in a network into groups or modules that exhibit higher internal connectivity and weaker connectivity between groups. Understanding network communities is essential for grasping how networks are structured and follow certain patterns. Moreover, these communities aid in visualizing and compressing networks. The significance of community structure lies in its connection to the hierarchical arrangement found in numerous complex real-world systems. This hierarchical setup aligns with the hierarchical community structure-networks consist of communities that encompass smaller ones, creating a cascade of diminishing size. Such a hierarchical arrangement optimizes the system's efficiency and functionality. Within this hierarchy, individual segments can integrate new technologies autonomously, thus lessening the potential for errors or failures propagating throughout the entire system (Shen 2013).

### Core-periphery structure

Core-periphery structure characterizes a complex network where nodes are divided into two main categories: the core and the periphery. The core nodes have stronger connections among themselves and typically exhibit higher levels of connectivity, while the periphery nodes have weaker connections and tend to be connected to the core nodes without strong connections between each other. This structure reflects a hierarchical arrangement where the core nodes play a central and influential role, while the periphery nodes have more limited interactions and often serve as conduits between the core elements. Core-periphery structures are common in various real-world networks, such as social networks, economic systems, and transportation networks (Holme 2005) (Csermely et al. 2013). According to Borgatti and Everett (2000), core-periphery structure in a network refers to the presence of a dense central core of nodes, which are more interconnected to each other than to the rest of the network, surrounded by a more sparsely connected periphery of nodes. The authors define the core-periphery structure as a pattern where certain group members are more densely related to one another than other members. They also note that the core-periphery structure can be observed at different levels of analysis, such as within a single network or between different networks.

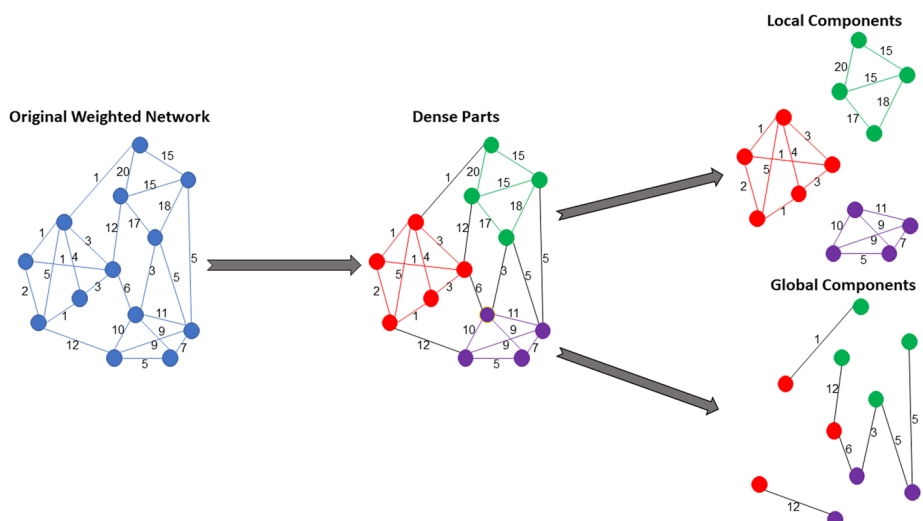

**Fig. 1** Component structure extraction of a toy example

## Component structure

The field of network science includes concepts such as community structure and core-periphery structure, which both are premised on the understanding that a network's connection density is not uniform. Dense areas within a network can form communities or core elements; the difference betwedense areas form the communities or the core elements of the network approach assumes that communities are sparsely connected, while the core-periphery structure sees peripheral nodes as poorly connected and to core nodes. The component structure combines these approaches, with dense areas forming local components tied together by proxy links and nodes to form global components. Identifying the global components is straightforward once the dense areas have been extracted. Definitions of dense areas used in community detection or multi-core-periphery studies can be used to extract local components (Diop et al. 2021).

*Toy example* Figure 1 illustrates a toy example of network decomposition into components. First, the dense parts are uncovered using algorithms designed to detect areas of high connectivity within the network. Nodes and links that belong to the same region share the same color. Inter-community links that connect the different dense parts are colored in black. The toy network contains three dense parts, which are colored in green, red, and yellow. We remove the inter-community links (links in black) to extract the local components. Each dense area represents a local component that possesses local information. We obtain the global component by eliminating the intra-community links and the isolated nodes. Global components play a significant role in information transmission among the network's different dense areas.

## Backbone extraction techniques

Backbone extraction techniques play a crucial role in network analysis as they seek to uncover the key structure and significant elements within intricate networks. By employing these methods, network complexity is effectively minimized while still retaining crucial connectivity patterns and capturing essential information about the network's organization and functionality. In this study, we utilize two distinct approaches to extract network backbones: the Global Threshold Method and the Disparity Filter. These techniques offer complementary perspectives by utilizing global and local measures to identify the most significant edges in a network. By employing both techniques, our objective is to assess the effectiveness of the proposed multilevel backbone framework, comparing its performance in local and global filtering processes. Through this analysis, we aim to determine whether a localized or broader global filtering approach yields superior results in extracting the network backbone. This evaluation will shed light on the optimal method for capturing the essential structural components of the network and facilitate a deeper understanding of its dynamics and underlying relationships.

### Global threshold

The Global Threshold Method, focusing on global measures, is a widely used and straightforward technique for network backbone extraction. This method involves setting a predefined threshold value and retaining only the edges whose weights exceed this threshold (Dai et al. 2018). The threshold can be defined in various ways, such as an absolute value, a proportion of the maximum observed edge weight, or the mean weight (Neal 2013). Global Threshold is popular due to its efficiency in producing sparser networks. However, it is important to note that Global Threshold may encounter limitations when applied to real-world networks. Many real-world networks exhibit uneven edge weights across multiple scales, introducing challenges. These include arbitrariness in threshold selection, structural bias in edge retention, and the assumption of a single scalar threshold (Neal 2014). These limitations impact the effectiveness and representativeness of the resulting network backbone.

### Disparity filter

The Disparity Filter adopts a local measures-based approach, focusing on the statistical significance of individual edge weights, to extract the network backbone. It is a widely used method that operates under the assumption that the weights of a node's edges, once normalized, conform to a uniform distribution. By comparing the normalized edge weights to this baseline model, it becomes possible to selectively filter out edges based on a desired significance level $\alpha$. The resulting filtered network, known as the backbone, retains only those statistically significant edges that conform to the null model. Importantly, an edge weight's significance can vary between nodes, as the method defines a distinct null model for each node, leading to potential asymmetry in significance assessment (Serrano et al. 2009; Yassin et al. 2023a).

## Multilevel backbone extraction framework based on the component structure

### General scheme

In real-world networks, node clusters exhibit varying degrees of interaction and can be called multicores, groups, or communities. Neglecting this structure and treating all groups equally can be problematic, notably when the weights of links in these groups differ significantly. To overcome this issue, the backbone extraction technique should be tailored to a specific group of nodes rather than the entire network (Hmaida et al. 2023). The component structure is a helpful tool in achieving this goal, as it divides the network into local and global components. Local components, corresponding to communities or cores, represent the densely connected parts of the network. Conversely, global components comprise nodes and links connecting these communities or cores. Thus, by concentrating on both local and global components, it becomes possible to discern the network's relevant connections more precisely.

The proposed framework involves two major steps, as shown in Fig. 2. The first step requires uncovering the dense parts of the network using a community detection algorithm or a core-periphery detection algorithm. Thus, the component structure of the network is established by first removing the inter-community links to isolate local components, followed by eliminating intra-community edges to obtain global components. In the second step, the backbone extraction technique is applied to each component rather than the whole network. Finally, the extracted backbones of both the global and local components are combined to reveal the overall backbone of the network. This approach enables the backbone extraction technique to accommodate the diverse topologies of each component, ensuring adaptability to potential heterogeneities.

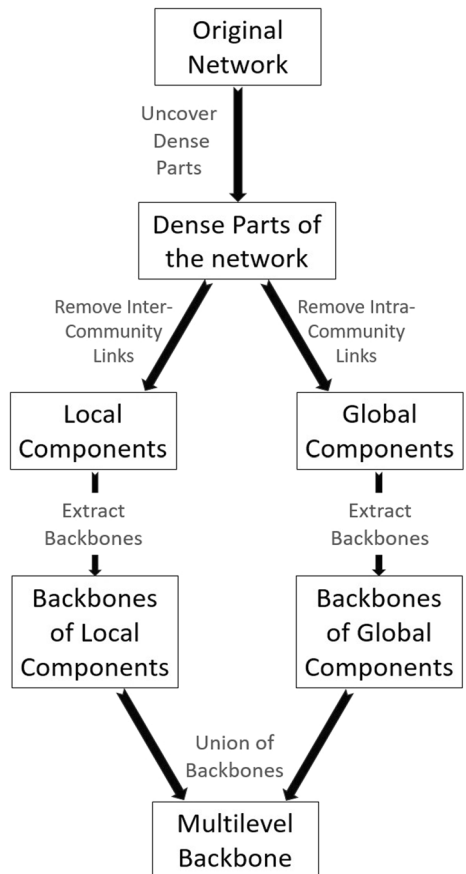

**Fig. 2** Scheme of the multilevel backbone extraction framework based on the component structure

**Multilevel backbone extraction algorithm**

In this subsection, we describe in detail the algorithm used for extracting the Multilevel backbone. Considering the following undirected and weighted network $G(V, E)$ , where $V = \{v_1, \ldots, v_N\}$ represents the set of nodes and $E = \{(v_i; v_j, \omega_{ij}) \backslash v_i, v_j \in V, \omega_{ij} \in \mathbb{R}\}$ denotes the set of edges with their weight. The algorithm (see Algorithm 1) employs several functions. The function $Extract\_dense(G)$ identifies the graph $G$'s dense parts. The function $Extract\_local(G)$ is used to extract the local components of the graph $G$ by removing the inter-community links. However, $Extract\_global(G)$ is the function that extracts the global components of the graph $G$ by removing the intra-community links. Finally, $Extract\_backbone(G)$ is the function that extracts the backbone of the graph $G$. This function can utilize any backbone extraction algorithm, tailored to the specific requirements of the network being analyzed.

**Algorithm 1** Multilevel backbone extraction algorithm

---

**Input** : Original graph $G(V, E)$
**Output**: Backbone graph $G_b(V_b, E_b)$

1   $G_d(V, E) \leftarrow Extract\_dense(G(V, E))$

2   $Local \leftarrow Extract\_local(G_d(V, E))$ //$Local = \{L_1, L_2, \ldots, L_n\}$, with $n$ is the total number of    local components.

3   $Global \leftarrow Extract\_global(G_d(V, E))$ //$Global = \{G_1, G_2, \ldots, G_m\}$, with $m$ is the total    number of global components.

4   $Local\_back \leftarrow \emptyset$ // Initialize empty local backbone set.

5   **for** $L_i \in Local$ **do**
6      $L\_back_i \leftarrow$ Extract_backbone$(L_i)$
7      $Local\_back \leftarrow Local\_back \cup L\_back_i$
8   **end**

9   $Global\_back \leftarrow \emptyset$ // Initialize an empty global backbone set.

10   **for** $G_i \in Global$ **do**
11      $G\_back_i \leftarrow$ Backbone$(G_i)$
12      $Global\_back \leftarrow Global\_back \cup G\_back_i$
13   **end**

14   $G_b \leftarrow Local\_back \cup Global\_back$ // Build the backbone as a union of local and global    backbones.

15   **return** $G_b(V_b, E_b)$

---

**Toy example**

Figure 3 uses a weighted and undirected toy network to illustrate the process of the proposed Multilevel backbone extraction algorithm. The original network is represented in the first graph. The Louvain algorithm is applied to the network to reveal its community structure. In the second graph, different colors highlight each community's nodes and edges, with inter-community links shown in black. To obtain the global components, intra-community links are removed, as shown in the graph on the right. Conversely, inter-community links are removed to isolate the local components, depicted in the graph on the left. The backbone's size is fixed at 30% of the network's edges. The Global Threshold algorithm is applied to each component to reveal the backbones. Finally, the Multilevel backbone is obtained by merging all the backbones of local and global components.

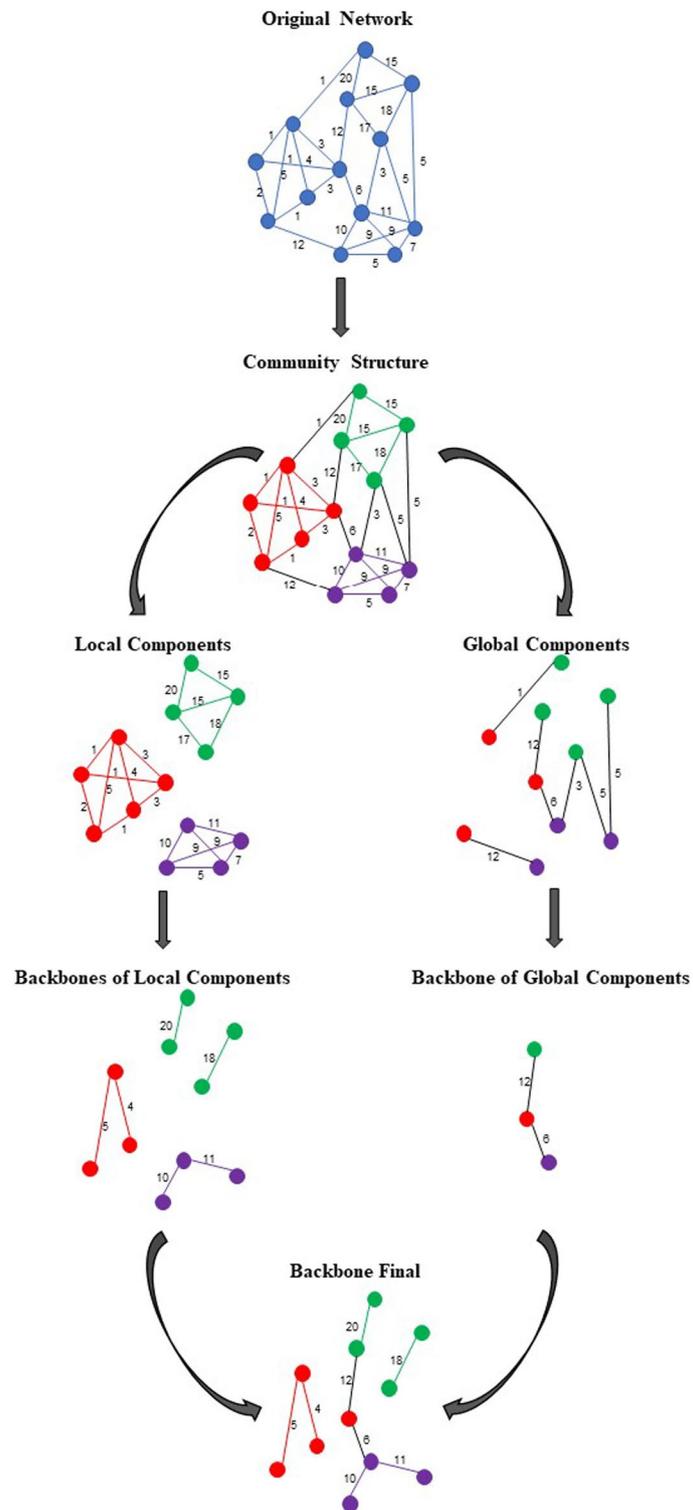

**Fig. 3** Multilevel backbone extraction of a toy example using global threshold

## Data and methods

### Data

In this section, we briefly outline the datasets utilized in our investigation. We selected networks from different domains (social, co-appearance, ecological, technological, transportation, and collaboration networks). To encompass a broad range of scenarios, we selected diverse networks with varying numbers of nodes and edges, ranging from hundreds to thousands. Small networks were chosen for qualitative evaluation of each backbone's behavior, while large networks were selected for quantitative analysis. The following paragraphs provide a brief overview of these networks. Table 1 reports the fundamental topological characteristics of these networks.

*Zachary's karate club* Friendship network among students in a karate club in 1970 at a US University. This network contains 33 nodes that signify the club's members and 77 edges that reflect their relationship. The weights assigned to these edges represent the strength of the relationships between the club members, with a number between 0 and 8 indicating the strength of their friendship (Ghalmane et al. 2020b).

*Wind surfers* Network of interpersonal contacts among windsurfers in the fall of 1986 in southern California. Nodes are windsurfers, edges represent the friendship between windsurfers, and edge weights indicate the surfers' interpersonal relationships (Almquist and Butts 2014).

*Madrid train bombing* Network of associations among the terrorists involved in the Madrid attack of March 11, 2004. Nodes denote terrorists, edges represent the connections between them, and edge weights represent how strongly they are connected (Rajeh et al. 2022).

*Les Misérables* The network of scene appearances of characters from "Les Misérables" by Victor Hugo. The nodes signify the personalities of this book, the edges represent the appearance of two nodes inside the same novel chapter, and the edge weights serve as a count of these occurrences (Coscia 2021).

*Unicode languages* This network describes the various nations and languages. Nodes represent languages and countries; edges represent the relationship between a country and the language in which it is spoken; and weights represent the percentage of the population that is literate in a particular language (Ghalmane et al. 2020b).

*World air transportation* Data about international flights was obtained from FlightAware. Since the data covers six days of 2018 (May 17−May 22), it ensures that less frequent

**Table 1** Fundamental topological characteristics of the real-world networks under study

| Network | V | E | $< k_w >$ | $\zeta$ | $\omega$ | $\epsilon$ | r | Q |
|---|---|---|---|---|---|---|---|---|
| Zachary's karate club | 33 | 78 | 13.59 | 0.256 | 0.139 | 0.492 | − 0.476 | 0.444 |
| Wind surfers | 43 | 336 | 56.09 | 0.564 | 0.372 | 0.679 | − 0.147 | 0.371 |
| Madrid train bombing | 62 | 243 | 8.81 | 0.561 | 0.121 | 0.448 | 0.029 | 0.435 |
| Les Misérables | 77 | 254 | 21.30 | 0.499 | 0.087 | 0.435 | − 0.165 | 0.565 |
| Unicode languages | 868 | 1255 | 0.697 | 0.00 | 0.003 | 0.255 | − 0.171 | 0.772 |
| World air transportation | 2734 | 16,665 | 131.49 | 0.257 | 0.004 | 0.283 | − 0.047 | 0.630 |
| Scientific collaboration | 16,726 | 47,594 | 9.23 | 0.360 | 0.0003 | 0.117 | 0.185 | 0.873 |

The total number of nodes is V. The number of edges is represented by E. $< k_w >$ signifies the average weighted degree. $\zeta$ indicates transitivity. $\omega$ symbolizes density. $\epsilon$ represents efficiency. The assortativity is r. Q stands for the network's weighted modularity connections are included. Airports are represented by nodes, while direct flights between airports are represented by edges. (Diop et al. 2021).

textitScientific collaboration Collaboration graphs for scientists. Nodes represent writers of papers in the arXiv category "Condensed Matter", the edges indicate co-authorship, and weights are the number of articles the authors have written together (Ghalmane et al. 2020b).

**Methods**

In this subsection, we evaluate our proposed method, wherein we utilize the component structure to extract the backbone. This evaluation is based on a variety of metrics. Firstly, we focus on fundamental global properties to assess the backbone's ability to preserve the global features of the original network. Secondly, we employ cumulative distributions and the Kolmogorov–Smirnov (KS) statistic to compare the microscopic characteristics of the extracted backbones with those of the original network. Additionally, we calculate the number of nodes and weights preserved in each extracted backbone to compare their mesoscopic properties. Lastly, we employ the Portrait Divergence Distance, Laplacian Spectrum Distance, and Network Laplacian Spectral Distance to measure the similarity between each backbone and the original network, thus determining the closest match.

### Basic global properties

textitDensity The density of networks is a measure of the number of links or connections between the nodes of a network relative to the total number of possible links in the network (Meunier et al. 2009).

For an undirected network with $V$ nodes and $E$ links, the density $\omega$ is given by:

$$\omega = \frac{2E}{V * (V - 1)} \tag{1}$$

textitDiameter The diameter of a network measures the maximum shortest path between any two nodes in the network. In other words, it is the longest path that must be traversed to go from one node to another, whereas "shortest" refers to the path with the fewest edges (Ng and Efstathiou 2006; Kaiser 2011). It's given by:

$$d = Max(d_{ij}) \tag{2}$$

where $d_{ij}$ represents the shortest path between nodes $i$ and $j$.

textitAverage shortest path The average shortest path is the number of edges found on all possible network node pairs' shortest paths (Jebabli et al. 2018). The average shortest path of a network is a measure of the typical distance between pairs of nodes in the network, where "distance" refers to the number of edges that must be traversed to go from one node to another using the shortest path (Watts et al. 1998). The average shortest path is defined as Dorogovtsev and Mendes (2004):

$$L = \frac{ln V}{ln \langle k \rangle} \tag{3}$$

where $V$ is the total number of nodes and $\langle k \rangle$ is the average degree.

*textitAverage node degree* The average node degree is the average number of edges connected to each node in the graph. In other words, it is the total number of edges in the graph divided by the total number of nodes in the graph (Meghanathan 2014). The formula for the average node degree is:

$$\langle k \rangle = \frac{\sum_i k_i}{V} \tag{4}$$

where $k_i$ represents the degree of node $i$.

*textitMax node degree* The max node degree is the highest number of edges connected to any single node in the graph (Meghanathan 2014). In other words, it is the maximum degree of any node in the graph given by:

$$k_{max} = Max(k_i) \tag{5}$$

where $k_i$ represents the degree of the node $i$.

*textitAssortativity coefficient* The assortativity coefficient measures the degree to which nodes in the graph tend to be connected to other nodes with a similar degree. It measures the correlation between the degrees of connected nodes in the graph (Newman 2003, 2002). The assortativity coefficient, denoted by $r$, is calculated as follows:

$$r = \frac{\sum_i (k_i * k_{nn_i}) - [\sum_i (0.5 * (k_i + k_{nn_i}))]^2}{\sum_i (0.5 * (k_i^2 + k_{nn_i}^2)) - [\sum_i (0.5 * (k_i + k_{nn_i}))]^2} \tag{6}$$

where $k_i$ denotes the degree of a node $i$ and $k_{nn_i}$ is the average degree of nodes connected to a node $i$.

The assortativity coefficient can take values between $-1$ and $1$. A positive value indicates that nodes tend to be connected to other nodes with a similar degree, while a negative value indicates that nodes tend to be connected to nodes with different degrees (McCormack et al. 2013).

*textitClustering coefficient* The clustering coefficient indicates the degree to which network nodes tend to cluster together (Gupta et al. 2015). It measures the density of triangles in the graph, where a triangle is a set of three nodes mutually connected by edges (Artameeyanant et al. 2016). The clustering coefficient is defined as Dorogovtsev and Mendes (2004):

$$C \equiv \frac{\langle n_i \rangle}{\langle k_i (k_i - 1)/2 \rangle} = \frac{\sum_k P(k) \langle n(k) \rangle}{(\langle k^2 \rangle - \langle k \rangle)/2} \tag{7}$$

where: $n_i$ represents the number of connections between the nearest neighbours of the node $i$, $k_i$ represents the degree of the node $i$, $P(k)$ represents the degree distribution, $n(k)$ is the number of vertices of degree $k$, and $\langle k \rangle$ is the average degree.

### Distributions

Two cumulative distributions are utilized to describe the microscopic characteristics of the networks. One is related to node degree, and the second is linked to edge weight.

textitCumulative degree distribution The cumulative degree distribution of a network is a measure of the number of nodes in the network that have a degree greater than or equal to a certain value. It is calculated by summing up the degrees of all nodes in the network that are greater than or equal to a certain degree and dividing by the total number of nodes in the network (Wang et al. 2021).

textitCumulative weight distribution However, the cumulative weight distribution of a network is a measure of the total weight of edges in the network that have a weight greater than or equal to a certain value. It is calculated by summing up the weights of all edges in the network that are greater than or equal to a certain weight and dividing by the total weight of all edges in the network.

textitKolmogorov-Smirnov statistic The Kolmogorov-Smirnov (KS) statistic is a nonparametric test that compares the cumulative distribution functions of two probability distributions. It quantifies the maximum vertical distance (or supremum) between the cumulative distribution functions of the two distributions being compared (Goldstein et al. 2004). The Kolmogorov-Smirnov statistic ($K$) is calculated as follows:

$$K = sup_x |F(x) - S(x)| \tag{8}$$

Where $sup_x$ represents the maximum value over all $x$, and $F(x)$ and $G(x)$ are the cumulative distribution functions of the two distributions.

### Distances

textitPortrait divergence distance Portrait divergence distance is a measure of the similarity between two networks. It is based on graph portraits, which are compact representations of the local connectivity patterns in a network (Lafhel et al. 2021). The graph portraits of the two networks are first computed to compute the portrait divergence distance between two networks. Then, the distance between the two portraits is calculated using a suitable distance metric, such as the Euclidean or Manhattan distance. The portrait divergence distance captures the differences between the local connectivity patterns in the two networks, rather than the differences in global network properties such as degree distribution or clustering coefficient. It is particularly useful for comparing networks with similar global properties but differ in their local connectivity patterns (Bagrow and Bollt 2019).

textitLaplacian spectrum distance The Laplacian spectrum distance measures the similarity between two graphs based on the eigenvalues of their Laplacian matrices. The Laplacian matrix of a graph is a matrix that encodes the local connectivity structure of the graph. To compute the Laplacian spectrum distance between two graphs, the Laplacian spectra of the two graphs are first computed, which is a vector of the eigenvalues of their Laplacian matrices. Then, the distance between the two spectra is calculated using a suitable distance metric, such as the Euclidean distance or the spectral angle distance (Grone et al. 1990). The Laplacian spectrum distance captures the structural similarities and differences between the two graphs, and it is particularly useful for comparing graphs that have similar degree distributions but different connectivity patterns (Mohar et al. 1991).

textitNetwork Laplacian spectral distance The Network Laplacian Spectral Distance (NetLSD) measures the distance between two networks based on their Laplacian spectral properties. To compute the NetLSD between two networks, the Laplacian matrix of the networks is first constructed. Then, the eigenvalues of the Laplacian matrices are computed and sorted in non-decreasing order. Finally, the distance between the two sets of eigenvalues is calculated using a suitable distance metric, such as the Euclidean distance or the spectral angle distance (Tsitsulin et al. 2018). The NetLSD is generally considered a more sophisticated measure than the Laplacian Spectral Distance because it considers more information about the structure of the networks being compared. The Laplacian Spectral Distance only considers the eigenvalues of the Laplacian matrices and does not consider the eigenvectors. While the Network Laplacian Spectral Distance considers both the eigenvalues and the eigenvectors of the Laplacian matrices (Tantardini et al. 2019).

### Mesoscopic properties

textitQuantitative analysis At the mesoscopic level, we use two indicators to quantify the proposed multilevel backbone.

- Fraction of preserved nodes in the backbone,
- Fraction of preserved weight in the backbone,

textitModularity Modularity is a fundamental concept in network analysis, particularly in studying complex networks. It quantifies the strength of dividing a network into modules, or communities. High modularity indicates a clear division into communities, reflecting a network structure where nodes are more likely to connect within their group than with nodes in other groups. Modularity is a key measure for understanding the community structure in networks. Identifying the presence and extent of community-like structures in various types of networks, including social, biological, and technological networks is crucial. High modularity often suggests functional or organizational segmentation within the network (Newman 2006).

Modularity is typically calculated using Newman's modularity measure (Newman 2006), defined as:

$$Q = \frac{1}{2m} \sum_{ij} \left[ A_{ij} - \frac{k_i k_j}{2m} \right] \delta(c_i, c_j) \tag{9}$$

where $A_{ij}$ represents the edge weight between nodes $i$ and $j$, $k_i$ and $k_j$ are the degrees of nodes $i$ and $j$, $m$ is the total weight of all edges in the network, $c_i$ and $c_j$ are the communities of nodes $i$ and $j$, and $\delta$ is the Kronecker delta function.

textitInter-community connectivity Inter-Community Connectivity pertains to the connections between distinct communities or modules within a network. This property is pivotal in understanding how different groups or communities within a network interact and communicate with each other. Inter-Community Connectivity is instrumental in revealing a network's integrative and segregative properties. It highlights the pathways through which information, resources, or influence flow between different communities, thereby elucidating the overall network structure and function (Girvan and Newman 2002). The Inter-Community Connectivity $ICC_{total}$ is given by:

$$ICC_{total} = \sum_{(i,j) \in E_{inter}} w_{ij} \tag{10}$$

Where $E_{inter}$ is the set of edges that connect nodes in different communities and $w_{ij}$ is the weight of the edge between nodes $i$ and $j$.

High inter-community connectivity can indicate a network with integrated communities, whereas low connectivity may suggest segregated or isolated communities.

textitIntra-community connectivity Intra-Community Connectivity refers to the extent and strength of connections within individual communities or modules of a network. This property is essential for understanding the internal cohesion and structural integrity of communities, which are groups of nodes more densely connected than nodes in other communities (Newman 2006). Intra-Community Connectivity provides insights into how nodes within a community are interconnected, reflecting the community's robustness and resilience to disruptions. It's defined as:

$$ICC'_{total} = \sum_{c \in C} \left( \sum_{(i,j) \in E_c} w_{ij} \right) \tag{11}$$

Where $C$ is the set of all communities, $E_c$ is the set of edges within a community $c$, and $w_{ij}$ is the weight of an edge between nodes $i$ and $j$.

High intra-community connectivity often indicates a strong, tightly knit community, which can be critical in networks where community structure dictates functional properties (Fortunato 2010).

## Experimental results and discussion

In this section, we conduct a thorough experimental analysis featuring real-world networks collected from various fields, each containing hundreds to thousands of nodes and edges to accommodate a range of circumstances. We analyze and compare the proposed Multilevel backbone with the Classical version of the backbone, where the backbone is directly extracted from the original graph using a backbone extraction algorithm. For Multilevel backbone extraction, we employ Louvain, the most efficient and effective community detection algorithm for large-scale networks (Blondel et al. 2008). Local components represent the communities, and global components consist of all the nodes and edges connecting these local components. In the two parts of the experiment, we choose to use global and local techniques for the backbone extraction of each component, which are the Global Threshold and the Disparity Filter. Finally, we construct the Multilevel backbone by merging all of these backbones.

For all the networks under study, we perform the same comparative analysis, ensuring that the generated backbones' size is consistently 30% of the original network size. This fixed percentage was chosen based on extensive empirical testing, which demonstrated that 30% provides an optimal balance between reducing network complexity and preserving key structural properties. Maintaining this consistent parameter setting ensures a fair comparison across different networks.

Firstly, we use two small networks, the Wind Surfers Network and the Madrid Train Bombing Network, to visualize the network and its extracted backbones. For the evaluation metrics, we present and discuss the results of all the networks in terms of Global properties and KS statistics. However, the Unicode Languages Network serves as an example in visualizing the degree and weight distribution, mesoscopic properties, and distance measures. In the Supplementary Materials (SM), a similar evaluation of the other networks is presented.

**Global threshold**

In this section, we employ the Global Threshold algorithm to simplify the network representation by retaining only the most significant edges based on a predefined threshold, allowing for efficient extraction of a sparser network backbone. However, acknowledging the limitations of the Global Threshold method in networks with variable density structures, we have introduced an adaptive thresholding mechanism that varies across different parts of the network. This mechanism calculates local thresholds based on the statistical properties of edges within each community structure, ensuring the retention of significant edges in both dense and sparse areas. Furthermore, our multiscale backbone strategy specifically addresses these limitations by extracting backbones from different dense parts of the network as well as from the global component. This approach ensures that important substructures are preserved and that the overall network complexity is accurately represented. These combined approaches enhance the robustness and accuracy of the backbone extraction process. Quantifying the effectiveness of the Multilevel backbone in revealing or retaining the organization of networks' structure beyond visual examination is valuable. Despite the significance of network visualization in network analytics, particularly for small- to medium-sized networks, it becomes increasingly challenging as the network size grows. To address this, we utilize several small networks, such as the Wind Surfers and Madrid Train Bombing networks, to visualize the differences between the Multilevel and Classical Global Threshold methods.

Figure 4 represents the Wind Surfers network alongside its Multilevel Global Threshold and Classical Global Threshold backbones, with a fraction of preserved edges fixed at 30%. The Multilevel backbone operates as a multi-threshold method, tailored to the weight distribution of each component rather than the entire network's weight distribution. In contrast, the Classical version eliminates the global component, leading to only one preserved link shared between nodes 1 and 7. Additionally, the Madrid Train Bombing network, containing six local components, is visualized in Fig. 5. Notably, the Classical Global Threshold removes the blue-colored local component, while the Multilevel Global Threshold preserves the overall component structure of the network, except for the small pink-colored local component, which consists of only two nodes and one link. The results demonstrate that the Multilevel Global Threshold employs multiple strategies adapted to the weight distribution of individual components rather than the overall network weight distribution.

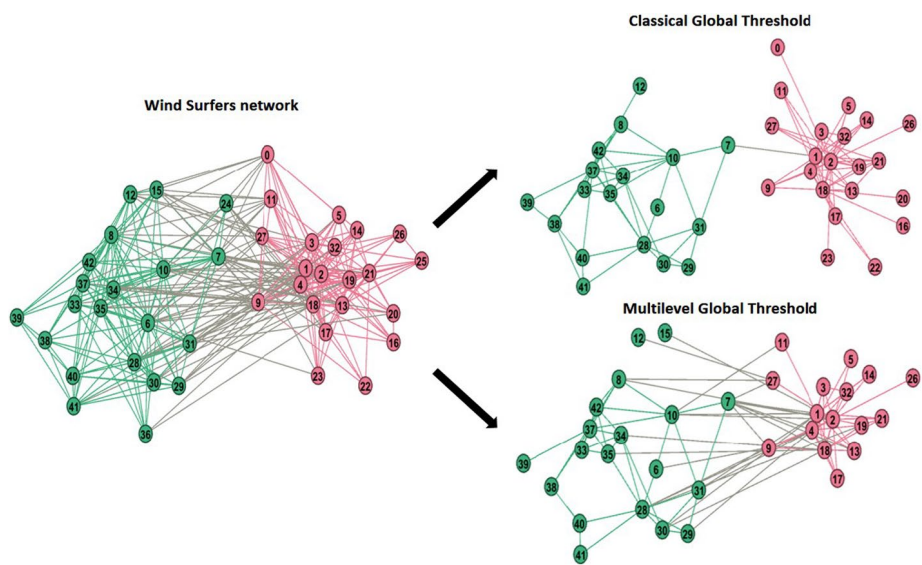

**Fig. 4** The extracted backbones using Global Threshold of the 'Wind Surfers' network (43 nodes and 336 edges), with a fraction of edges fixed at 30%

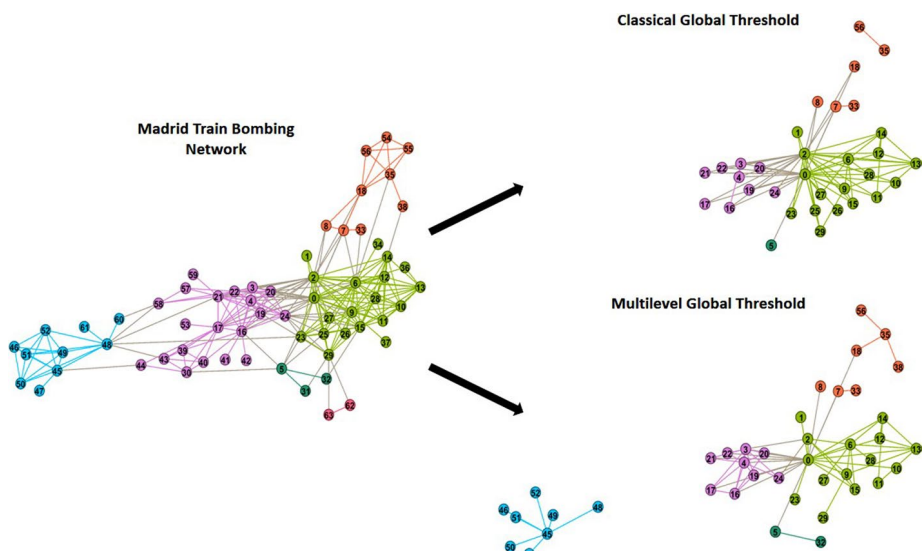

**Fig. 5** The extracted backbones using Global Threshold of the 'Madrid Train Bombing' network (62 nodes and 243 edges), with a fraction of edges fixed at 30%

### *Comparing the basic global properties*

Table 2 provides a comprehensive description of the global properties of the original networks under investigation, along with their corresponding extracted Classical and Multilevel Global Threshold backbones. For Zachary's Karate Club network, the Multilevel Global Threshold almost preserves the diameter, average shortest path, and assortativity, while the Classical Global Threshold closely matches the density value of the original network. Both backbones exhibit equivalent values for average node degree, max node degree, and clustering coefficient. Regarding the Wind Surfers network, the Multilevel

**Table 2** Global properties of the real-world networks under study and their Classical Global Threshold and Multilevel Global Threshold

| | V | E | ω | d | L | ⟨k⟩ | $k_{max}$ | r | C |
|---|---|---|---|---|---|---|---|---|---|
| Zachary's karate club | 34 | 78 | 1.39E−01 | 5 | 2.41 | 4.59 | 17 | −0.48 | 0.57 |
| Classical global threshold | 19 | 23 | 1.35E−01 | 7 | 3.23 | 2.42 | 6 | −0.19 | 0.21 |
| Multilevel global threshold | 23 | 23 | 1.21E−01 | 4 | 2.33 | 2.30 | 5 | −0.27 | 0.22 |
| Wind surfers | 43 | 336 | 3.72E−01 | 3 | 1.67 | 15.63 | 31 | −0.15 | 0.65 |
| Classical global threshold | 39 | 101 | 1.36E−01 | 5 | 2.61 | 5.18 | 21 | −0.30 | 0.54 |
| Multilevel global threshold | 36 | 101 | 1.60E−01 | 5 | 2.29 | 5.61 | 21 | −0.31 | 0.51 |
| Madrid train bombing | 64 | 243 | 1.21E−01 | 6 | 2.69 | 7.59 | 29 | 0.03 | 0.62 |
| Classical global threshold | 33 | 73 | 1.38E−01 | 3 | 1.90 | 4.42 | 29 | −0.65 | 0.73 |
| Multilevel global threshold | 41 | 72 | 8.87E−02 | 5 | 2.44 | 3.51 | 23 | −0.30 | 0.33 |
| Les Misérables | 77 | 254 | 8.68E−02 | 5 | 2.64 | 6.60 | 36 | −0.17 | 0.57 |
| Classical global threshold | 35 | 76 | 1.28E−01 | 6 | 3.02 | 4.34 | 13 | −0.04 | 0.67 |
| Multilevel global threshold | 41 | 76 | 9.27E−02 | 7 | 3.27 | 3.71 | 15 | −0.12 | 0.48 |
| Unicode languages | 868 | 1255 | 3.34E−03 | 8 | 4.21 | 2.89 | 141 | −0.17 | 0 |
| Classical global threshold | 399 | 376 | 4.74E−03 | 11 | 4.07 | 1.88 | 106 | −0.24 | 0 |
| Multilevel global threshold | 341 | 373 | 6.43E−03 | 12 | 4.04 | 2.19 | 105 | −0.24 | 0 |
| World air transportation | 2518 | 16,313 | 5.15E−03 | 9 | 3.62 | 12.96 | 242 | −0.07 | 0.47 |
| Classical global threshold | 802 | 4894 | 1.52E−02 | 7 | 3.14 | 12.20 | 154 | −0.17 | 0.39 |
| Multilevel global threshold | 918 | 4892 | 1.16E−02 | 7 | 3.40 | 10.66 | 119 | −0.07 | 0.39 |
| Scientific collaboration | 16,264 | 47,594 | 3.60E−04 | 18 | 6.63 | 5.85 | 107 | 0.18 | 0.64 |
| Classical global threshold | 9815 | 14,278 | 2.96E−04 | 25 | 8.09 | 2.91 | 44 | 0.04 | 0.33 |
| Multilevel global threshold | 9345 | 14,217 | 3.56E−04 | 20 | 8.16 | 3.04 | 40 | 0.08 | 0.34 |

*V* is the number of total nodes. *E* is the number of edges. *ω* denotes density. *d* is the diameter. *L* represents the average shortest path. ⟨*k*⟩ is the average node degree. $k_{max}$ represents the max node degree. *r* is the assortativity coefficient. *C* denotes the clustering coefficient and Classical Global Threshold yield similar values for all the global properties. In the case of the Madrid Train Bombing network, the Multilevel backbone demonstrates similar values for diameter and average shortest path as the original network. However, the Classical backbone better preserves the density, average node degree, max node degree, and clustering coefficient. Concerning the Les Misérables network, the Multilevel Global Threshold comes close to preserving global properties concerning density, assortativity, and clustering coefficient. The Classical version shows similarity to the original network in terms of diameter and average node degree. Meanwhile, the average shortest path and max node degree are equivalent for both Multiscale and Classical backbones. As for the Unicode Languages network, the Multilevel Global Threshold preserves only the average node degree, whereas the Classical version retains only the density. For the other properties, both backbones yield similar results. In the World Air Transportation network, the Multilevel Global Threshold preserves the average shortest path and assortativity, while the Classical Global Threshold maintains the average node degree of the original network. The remaining properties have comparable results for both backbones. In the Scientific Collaboration network, the Classical backbone does not preserve any properties, whereas the Multilevel backbone effectively preserves the density, diameter, average node degree, and assortativity of the original network. Additionally, the average shortest path, max node degree, and clustering coefficient results are similar for both the Classical and Multilevel backbones. Overall, one can say that the Multilevel Global Threshold can preserve the global properties of the original network, especially for the Scientific Collaboration network.

***Comparing the distributions***

textitCumulative degree distribution Let's now discuss the cumulative degree distribution presented in Fig. 6 for the Unicode Languages network and its extracted backbones, which are the Multilevel and the Classical Global Threshold. Interestingly, the Multilevel Global Threshold provides a sub-graph with a distribution slightly more similar to the original network than the Classical Global Threshold. Moreover, it is more likely to preserve the main characteristics of the original network than the Classical version. This finding is reinforced by the Kolmogorov-Smirnov (KS) statistics, which are highlighted in Table 3. Notably, the KS statistic is small (0.086), indicating a considerable resemblance between the Multilevel Global Threshold degree distribution and that of the original network. In contrast, there is a more substantial difference between the Classical Global Threshold degree distribution and the original network degree distribution.

Similar results are observed for the other networks under examination (see Supplemental Materials). The cumulative degree distribution of the overall networks and their backbones exhibits analogous behavior to that of the Unicode Languages network. Except for Zachary's karate club and Les Misérables networks, the KS statistic of the Classical Global Threshold is consistently lower than that of the Multilevel Global Threshold, as depicted in Table 3. This signifies that the cumulative degree distribution of the Classical backbone is more closely aligned with the original network compared to the Multilevel version.

textitCumulative weight distribution In analyzing the cumulative weight distribution (Fig. 7), we observe that the Multilevel Global Threshold preserves nearly all scales, with only the area of extremely light weights being impacted. Conversely, the Classical Global Threshold eliminates a significant portion of small weights, retaining only high-scale ones. This finding is further supported by the results of the KS statistic, which confirm that the Multilevel Global Threshold weight distribution is more similar to the original network weight distribution compared to the Classical Global Threshold weight distribution.

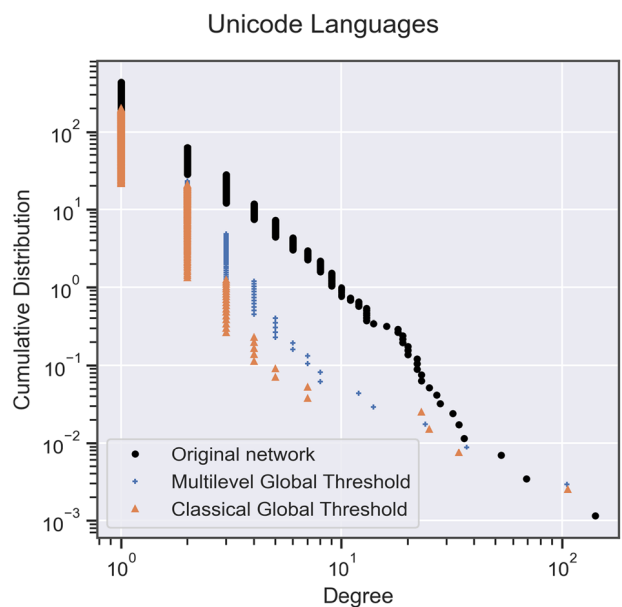

**Fig. 6** Cumulative degree distribution for the unicode language network and its extracted multilevel and classical global threshold

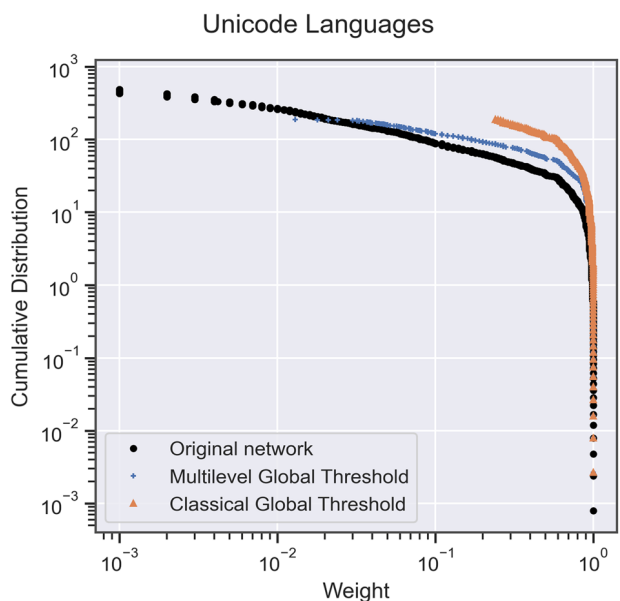

**Fig. 7** Cumulative weight distribution for the unicode language network and its extracted multilevel and classical global threshold

**Table 3** KS statistics between the extracted backbone degree distribution and the original network degree distribution, as well as between the extracted backbone weight distribution and the original network weight distribution

| Network | Backbone | KS (Degree Distribution) | KS (Weight Distribution) |
|---|---|---|---|
| Zachary's karate club | Classical global threshold | 0.339 | 0.644 |
| | Multilevel global threshold | 0.371 | 0.557 |
| Wind surfers | Classical global threshold | 0.797 | 0.652 |
| | Multilevel global threshold | 0.699 | 0.497 |
| Madrid train bombing | Classical global threshold | 0.443 | 0.278 |
| | Multilevel global threshold | 0.381 | 0.283 |
| Les Misérables | Classical global threshold | 0.296 | 0.664 |
| | Multilevel global threshold | 0.337 | 0.532 |
| Unicode languages | Classical global threshold | 0.176 | 0.697 |
| | Multilevel global threshold | 0.086 | 0.481 |
| World air transportation | Classical global threshold | 0.088 | 0.700 |
| | Multilevel global threshold | 0.049 | 0.550 |
| Scientific collaboration | Classical global threshold | 0.321 | 0.618 |
| | Multilevel global threshold | 0.282 | 0.510 |

Similar patterns have been observed in other networks under investigation (refer to the Supplemental Materials). Most networks and their backbones exhibit behavior similar to the cumulative weight distribution of the Unicode Languages network. However, in the case of the Madrid Train Bombing network, as indicated in Table 3, the KS statistic between the network and its Classical Global Threshold is slightly lower than the KS statistic between the network and its Multilevel Global Threshold. Therefore, in contrast to the Multilevel backbone, the cumulative weight distribution of the Classical Global Threshold is slightly more similar to the original network.

***Comparing the distances***

Let's now examine different distance measures. Figure 8 illustrates the progression of Portrait Divergence Distance (left panel), Laplacian Spectrum Distance (middle panel), and Network Laplacian Spectral Distance (right panel) between the Unicode Languages network and its extracted backbones. In the left panel, we observe that, for all link fractions below 90%, the Multilevel Global Threshold exhibits a lower distance than the Classical Global Threshold, indicating a closer similarity between the Multilevel Global Threshold and the original network in terms of visual characteristics. Concerning the Laplacian Spectrum Distance, the Multilevel Global Threshold is closer to the original network than the Classical backbone for link fractions less than 60%. The right panel of the Figure further confirms that for percentages below 25% and above 55%, the Network Laplacian Spectral Distance (NetLSD) value for the Multilevel Global Threshold is smaller than the Classical Global Threshold values. In fact, by comparing the underlying network structures, we can conclude that the Multilevel Global Threshold is closer to the original network than the Classical Global Threshold. Overall, the collective distance measurements support the notion that the Multilevel Framework, employing a Global Threshold algorithm, preserves more properties of the original graph than the Classical version.

Discrepancies between this network and the others can be observed (refer to the Supplementary Materials). Specifically, concerning Zachary's karate club network, when the fraction of preserved edges exceeds 70%, the Laplacian Spectrum Distance between the original network and the Classical Global Threshold is smaller than the distance between the original network and the Multilevel Global Threshold. Additionally, for percentages below 20% and above 40%, the Classical backbone shows closer proximity to the original

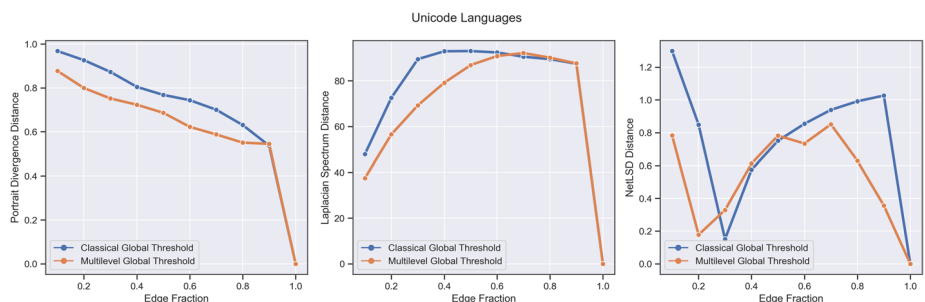

**Fig. 8** Distance measurements according to the fraction of preserved links between the original network and the Multilevel Global Threshold (curve in orange) and between the original network and the Classical Global Threshold (curve in blue). Panels represent respectively Portrait Divergence Distance (Left), Laplacian Spectrum Distance (Middle), and Network Laplacian Spectral Distance (Right)

network compared to the Multilevel version. In the case of the Wind Surfers network, the Portrait Divergence Distance indicates that the Multilevel Global Threshold is closer to the original for fractions under 70%. However, the Laplacian Spectrum Distance reveals that the Multilevel Global Threshold significantly deviates from the original network for fractions of edges between 30% and 70%. Regarding the Les Misérables network, both the Portrait Divergence Distance and the Laplacian Spectrum Distance confirm that the Classical Global Threshold is the closer backbone to the original network. Nevertheless, concerning NetLSD, for fractions below 30%, the distance between the original network and the Multilevel Global Threshold is higher, implying that, for this fraction, the Classical Global Threshold tends to be closer to the original network. Lastly, for the World air transportation network, we can deduce that after a certain fraction of preserved links, the Classical Global Threshold tends to be closer to the original network than the Multilevel Global Threshold.

### Comparing the mesoscopic properties

textitQuantitative analysis As we apply the Multilevel Global Threshold and Classical Global Threshold to extract the backbone of the network, we present statistics regarding the relative sizes of each backbone in terms of fractions of total weight and nodes preserved while varying the fraction of retained edges, as shown in Fig. 9. Specifically, for the Unicode Languages network, the Multilevel Global Threshold reduces the number of nodes more than the Classical Global Threshold. Looking at the right panel, we also observe that the Classical Global Threshold retains a higher proportion of weight than the Multilevel Global Threshold. This difference can be attributed to the Multilevel Global Threshold's interest in both high and low weights, while the Classical backbone filters out low weights and preserves only the high weights of the network. Similarly, the other networks (refer to the Supplemental Materials) exhibit similar behavior regarding the preserved weight fraction. However, for networks such as Zachary's karate club, Les Misérables, and World Air Transportation, the Multilevel Global Threshold preserves a higher fraction of nodes than the Classical version for all link fractions below 60%.

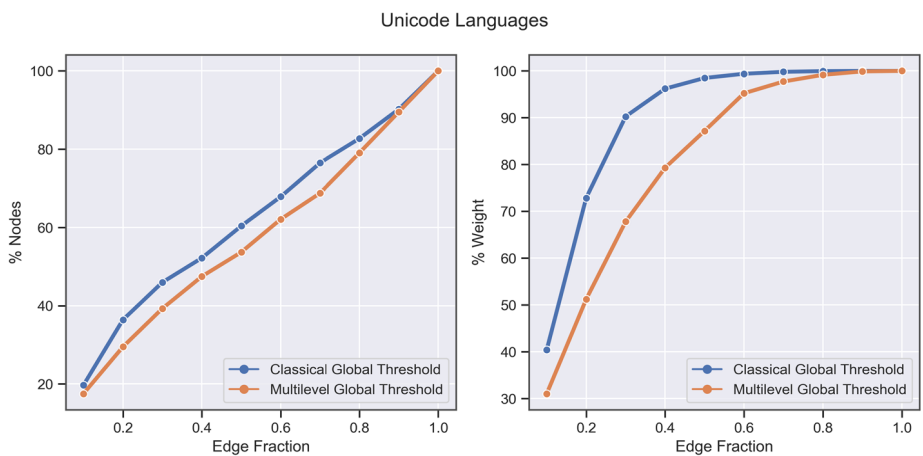

**Fig. 9** Fraction of preserved nodes and weights as a function of the fraction of edges maintained by the Multilevel and Classical Global Threshold of the Unicode Language network

**Table 4** Modularity of the real-world networks under study and their Classical Global Threshold and Multilevel Global Threshold

| Network | Modularity | | |
|---|---|---|---|
| | Original network | Classical global threshold | Multilevel global threshold |
| Zachary's karate club | 0.443 | 0.442 | 0.480 |
| Wind surfers | 0.371 | 0.408 | 0.375 |
| Madrid train bombing | 0.435 | 0.061 | 0.386 |
| Les Misérables | 0.565 | 0.522 | 0.523 |
| Unicode languages | 0.771 | 0.789 | 0.733 |
| World air transportation | 0.630 | 0.388 | 0.441 |
| Scientific collaboration | 0.873 | 0.903 | 0.883 |

textitModularity The modularity of the examined real-world networks and their extracted backbones using Global Threshold is shown in Table 4. The essence of modularity as an evaluative metric lies in its ability to quantify the strength of the division within a network into communities. Consequently, higher modularity values are indicative of a pronounced community structure.

The Classical Global Threshold seems to either maintain or enhance modularity in most cases, indicating a potential for reinforcing the prE−existing community structures within the networks. Notably, this method significantly increased modularity in the Wind Surfers and Scientific Collaboration networks, suggesting it might be particularly effective in networks with strong inherent community structures. On the other hand, the Multilevel Global Threshold generally exhibits a more conservative approach. In the case of the Madrid Train Bombing network, the Multilevel Global Threshold better preserved the modularity of the original network compared to the Classical Global Threshold, which significantly diminished it. This suggests that the Multilevel Global Threshold may be more adept at maintaining the integrity of the original network's structure, particularly in cases where the Classical Global Threshold might over-prune connections and disrupt the community layout.

Overall, if the objective is to maintain modularity closest to the original network, the Multilevel Global Threshold appears to be the more consistent method across varied network structures. It strikes a balance between preserving the original network's modularity and extracting a meaningful backbone, suggesting its utility in applications where the integrity of the original network's community structure is paramount.

textitParticipation coefficient Figure 10 illustrates the distribution of the Participation Coefficient across the original Unicode Language network and its backbones, processed through the Classical Global Threshold and the Multilevel Global Threshold techniques. The Participation Coefficient measures the diversity of intermodular connections of nodes in a network. A higher Participation Coefficient suggests that a node is connected to many different communities, while a lower value indicates that a node's connections are largely confined within its community.

The plot indicates that the original network has a certain distribution of the Participation Coefficient, a reference for the effectiveness of the two thresholding methods. The

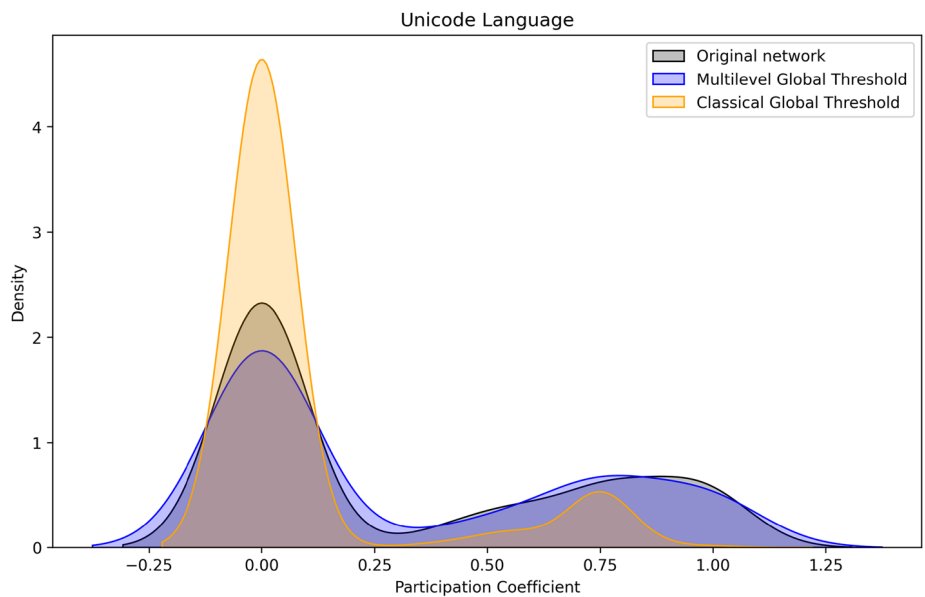

**Fig. 10** Participation coefficient for the unicode language network and its extracted multilevel and classical global threshold

Classical Global Threshold appears to skew the distribution towards lower Participation Coefficient values, suggesting a tendency to confine nodes within their communities, reducing the intermodular connectivity. On the other hand, the Multilevel Global Threshold shows a distribution that is more aligned with the original network's, maintaining a broader spread of Participation Coefficient values. This implies that the Multilevel Global Threshold preserves the diversity of intermodular connections better than the Classical Global Threshold.

From this visualization, we can infer that the Multilevel Global Threshold is more effective at preserving the network's community structure, as measured by the Participation Coefficient. The nodes maintain a level of connectivity to multiple communities that is more similar to the original network structure compared to the Classical Global Threshold, which seems to promote a higher level of modularity by restricting nodes to within-community connections.

This analysis suggests that the Multilevel Global Threshold would be the preferred method for maintaining network connections' structural diversity. It preserves the intricate balance of within-community and between-community connections that characterize the original network, which is crucial for understanding the network's functional dynamics.

*Inter-community connectivity* Table 5 provides values of the inter-community connectivity of networks under study and their backbones using Classical Global Threshold and Multilevel Global Threshold techniques. Inter-community connectivity is a crucial metric in network analysis, as it measures the extent of connections between distinct communities within a network. A higher value indicates more community connections, often associated with a robust and integrated network structure.

Upon examining the inter-community connectivity values, it is apparent that both thresholding techniques generally reduce this metric compared to the original network values, indicative of a loss of cross-community links post-thresholding. However, the degree to which each method preserves the original network's inter-community connectivity varies.

**Table 5** Inter-Community Connectivity of the real-world networks under study and their Classical Global Threshold and Multilevel Global Threshold

| Network | Inter-community connectivity | | |
|---|---|---|---|
| | Original network | Classical global threshold | Multilevel global threshold |
| Zachary's karate club | 51.00 | 21.00 | 19.00 |
| Wind surfers | 118.00 | 20.00 | 59.00 |
| Madrid train bombing | 55.00 | 27.00 | 17.00 |
| Les Misérables | 154.00 | 101.00 | 101.00 |
| Unicode languages | 37.53 | 27.43 | 35.70 |
| World air transportation | 1041054 | 945,906 | 803,281 |
| Scientific collaboration | 3001.50 | 1488.50 | 1747.51 |

For instance, in Zachary's Karate Club network, the original inter-community connectivity is 51.00, which is reduced to 21.00 by the Classical Global Threshold and, to an even lower extent, by the Multilevel Global Threshold at 19.00. This pattern, where the Multilevel Global Threshold retains connectivity closer to the original structure, is also reflected in the Wind Surfers network, although the Classical Global Threshold results in a drastic reduction from the original 118.00 to 20.00, whereas the Multilevel Global Threshold maintains higher connectivity at 59.00. The Madrid Train Bombing network offers a contrasting result, with the Classical Global Threshold preserving more of the original network's connectivity (27.00) than the Multilevel Global Threshold (17.00). The same behaviour is seen in the World Air Transportation Network; the Multilevel backbone preserves less than the Classical backbone regarding inter-community connectivity. Les Misérables, on the other hand, shows equal preservation by both methods, each maintaining the connectivity at 101.00, a decrease from the original value of 154.00. For the Unicode Languages network, the Multilevel Global Threshold preserves more connectivity (35.70) than the Classical Global Threshold (27.43), against an original value of 37.53. Finally, the Scientific Collaboration network, which starts with a significantly high inter-community connectivity of 3001.50, retains 1488.50 with the Classical Global Threshold and 1747.51 with the Multilevel Global Threshold, indicating that the Multilevel Global Threshold is more effective at preserving inter-community connections in this case.

In summary, our results suggest that the effectiveness of the Classical Global Threshold and the Multilevel Global Threshold in preserving the original network's inter-community connectivity is context-dependent. While the Multilevel Global Threshold better maintains connectivity for Wind Surfers and Unicode Languages networks, the Classical Global Threshold is more effective for the Zachary's Karate Club and Madrid Train Bombing networks. Both methods are equivalent for networks like Les Misérables, and for the highly interconnected Scientific Collaboration network, the Multilevel Global Threshold shows superior performance. This nuanced analysis indicates that careful consideration must be given to the backbone extraction method based on the specific network's structure and the analytical objectives of preserving inter-community connectivity.

textitIntra-community connectivity Table 6 illustrates the intra-community connectivity values of the networks under evaluation and their extracted backbones: the Classical Global Threshold and the Multilevel Global Threshold. Intra-community connectivity measures the density of links within individual communities, and high values typically reflect a tightly-knit community structure.

For Zachary's Karate Club, the original network exhibits an intra-community connectivity of 180.0, which is reduced to 81.00 by the Classical Global Threshold and further marginally by the Multilevel Global Threshold to 80.00. This marginal difference suggests that both methods reduce connectivity within communities for this network. In the Wind Surfers network, the original intra-community connectivity of 1088.00 is significantly reduced by both methods; however, the Classical Global Threshold maintains higher connectivity (839.00) compared to the Multilevel Global Threshold (766.00), indicating a better preservation of the original network's intra-community structure. The Madrid Train Bombing network sees an increase in intra-community connectivity from 85.00 to 94.00 with the Multilevel Global Threshold, contrary to a reduction to 85.00 by the Classical Global Threshold, positioning the Multilevel approach as a more conservative method in terms of preserving the network's original connectivity. In Les Misérables, the original intra-community connectivity of 666.00 is reduced to 429.00 by the Classical Global Threshold and, to a slightly lower extent, by the Multilevel Global Threshold of 412.00. Both methods decrease connectivity, with the Classical Threshold showing a closer value to the original network. The Unicode Languages network originally possesses an intra-community connectivity of 265.15, which is reduced to 245.75 by the Classical Global Threshold and further to 169.48 by the Multilevel Global Threshold, implying that the Classical method is closer to the original connectivity value. For World Air Transportation, the Classical Global Threshold and the Multilevel Global Threshold reduce the original intra-community connectivity; however, the Classical method preserves the higher connectivity, which shows a closer value to the original network than the Multilevel method. Lastly, the Scientific Collaboration network, with a high original connectivity of 24208.00, sees reductions to 17252.17 by the Classical Global Threshold and to 15860.66 by the Multilevel Global Threshold. Although both methods significantly reduce intra-community connectivity, the Classical Global Threshold retains a value closer to the original network.

**Table 6** Intra-Community Connectivity of the real-world networks under study and their Classical Global Threshold and Multilevel Global Threshold

| Network | Intra-Community Connectivity | | |
| --- | --- | --- | --- |
| | Original Network | Classical Global Threshold | Multilevel Global Threshold |
| Zachary's karate club | 180.00 | 81.00 | 80.00 |
| Wind surfers | 1088.00 | 839.00 | 766.00 |
| Madrid train bombing | 227.00 | 85.00 | 94.00 |
| Les Misérables | 666.00 | 429.00 | 412.00 |
| Unicode languages | 265.15 | 245.75 | 169.48 |
| World Air transportation | 3,475,074 | 2,642,841 | 2,634,522 |
| Scientific collaboration | 24208.00 | 17252.17 | 15860.66 |

In summary, applying the Classical Global Threshold and Multilevel Global Threshold has distinct impacts on the intra-community connectivity of networks. The Classical Global Threshold tends to preserve intra-community connectivity more closely to the original values for most networks, except for the Madrid Train Bombing network, where the Multilevel Global Threshold is more aligned with the original structure.

### Disparity filter

In this section, we employ the Disparity Filter as an algorithm for backbone extraction. The Disparity Filter captures statistically significant edges at the local level, considering the unique characteristics of individual nodes and their respective null models. This nuanced approach provides a deeper understanding of the network's structure, uncovering hidden patterns or relationships that global filtering methods might overlook. Although network visualization plays a vital role in analyzing smaller to medium-sized networks, it becomes increasingly challenging as the network grows in size. To highlight the distinctions between the Multilevel Disparity Filter and the Classical Disparity Filter, we present visualizations of two small networks and their extracted backbones: the Wind Surfers and the Madrid Train Bombing networks.

The Wind Surfers network and its Multilevel Disparity Filter and Classical Disparity Filter, with a fixed fraction of edges at 30%, are shown in Fig. 11. Remarkably, the Classical Disparity Filter tends to eliminate the global component, resulting in only three preserved edges shared between the two local components, which are (7,1), (7,2), and (7,9). In contrast, the Multilevel Disparity Filter successfully maintains all the network components. Furthermore, Fig. 12 displays the network of the Madrid Train Bombing and its extracted backbones, with the same fraction of preserved edges (30%). The Classical Disparity Filter filters out two local components (colored blue and hunter green). On the other hand, the Multilevel Disparity Filter preserves the network's component structure. Except for the small local component highlighted in pink, which contains only two nodes connected by one edge, the Multilevel and Classical backbones remove this component. In conclusion, the findings indicate that the Multilevel Disparity Filter functions as diverse approaches tailored to the weight distribution of individual components, rather than being dependent on the overall network weight distribution. Overall, after pruning 70% of the network, the results demonstrate that the Multilevel Disparity Filter operates as multiple strategies customized to each component's weight distribution rather than relying on the overall network weight distribution.

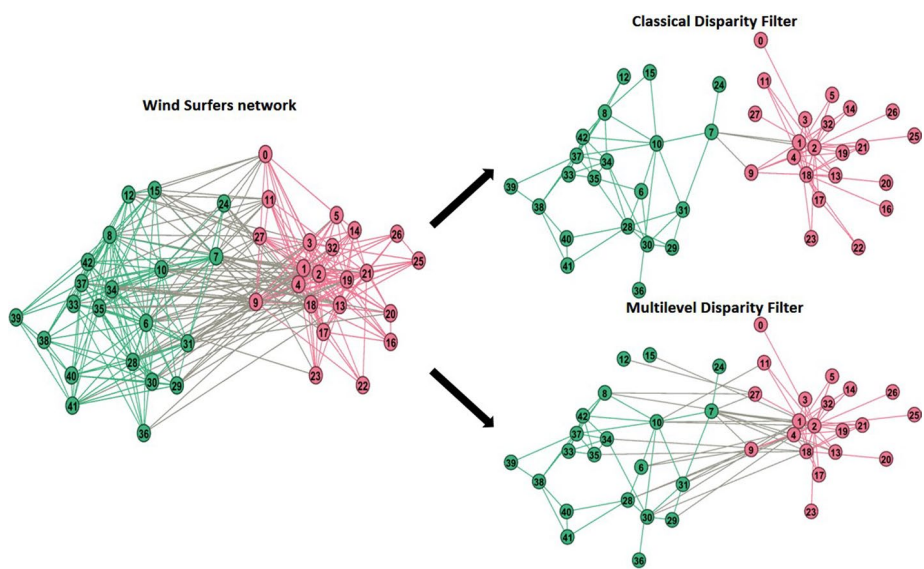

**Fig. 11** The extracted backbones using Disparity Filter of 'Wind Surfers' network (43 nodes and 336 edges), with a fraction of edges fixed at 30%

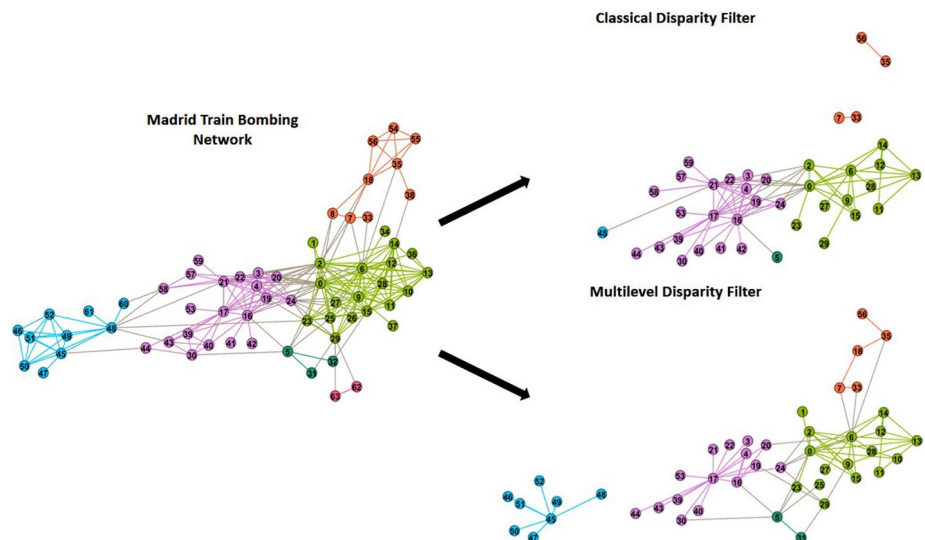

**Fig. 12** The extracted backbones using Disparity Filter of 'Madrid Train Bombing' (62 nodes and 243 edges), network with a fraction of edges fixed at 30%

***Comparing the basic global properties***

The global properties of the studied networks, their Classical Disparity Filter, and their Multilevel Disparity Filter backbones are presented in Table 7. The results for the Classical Disparity Filter and the Multilevel Disparity Filter exhibit some similarities. For the first network, Zachary's Karate Club, both backbones display nearly identical values for density, diameter, average shortest path, average and max node degree. However, there are slight differences in assortativity and clustering coefficients, where the Multilevel Disparity Filter's values are closer to those of the original network than the Classical version. In the case of the Wind Surfers network, the Multilevel Disparity Filter preserves the diameter, average shortest path, and max node degree of the original network better than the Classical version. Conversely, the Classical Disparity Filter has a clustering coefficient closer to the original network. The other global properties show similar values for

**Table 7** Global properties of real−world networks under study, their Classical Disparity Filter, and their Multilevel Disparity Filter

|  | V | E | ω | d | L | ⟨k⟩ | $k_{max}$ | r | C |
|---|---|---|---|---|---|---|---|---|---|
| Zachary's Karate Club | 34 | 78 | 1.39E−01 | 5 | 2.41 | 4.59 | 17 | −0.48 | 0.57 |
| Classical disparity filter | 21 | 23 | 1.10E−01 | 7 | 3.41 | 2.19 | 6 | −0.26 | 0.17 |
| Multilevel disparity filter | 21 | 23 | 1.10E−01 | 7 | 3.40 | 2.19 | 6 | −0.31 | 0.24 |
| Wind Surfers | 43 | 336 | 3.72E−01 | 3 | 1.67 | 15.63 | 31 | −0.15 | 0.65 |
| Classical disparity filter | 43 | 101 | 1.12E−01 | 6 | 3.10 | 4.70 | 19 | −0.39 | 0.54 |
| Multilevel disparity filter | 41 | 101 | 1.23E−01 | 4 | 2.37 | 4.93 | 24 | −0.32 | 0.46 |
| Madrid Train Bombing | 64 | 243 | 1.2E−01 | 6 | 2.69 | 7.59 | 29 | 0.03 | 0.62 |
| Classical disparity filter | 39 | 73 | 9.85E−02 | 4 | 2.43 | 3.74 | 16 | −0.46 | 0.37 |
| Multilevel disparity filter | 45 | 72 | 7.27E−02 | 5 | 2.82 | 3.20 | 14 | −0.46 | 0.28 |
| Les Misérables | 77 | 254 | 8.68E−02 | 5 | 2.64 | 6.60 | 36 | −0.17 | 0.57 |
| Classical disparity filter | 45 | 76 | 7.68E−02 | 7 | 3.17 | 3.38 | 11 | −0.04 | 0.36 |
| Multilevel disparity filter | 47 | 76 | 7.03E−02 | 6 | 3.07 | 3.23 | 13 | −0.006 | 0.43 |
| Unicode languages | 868 | 1255 | 3.34E−03 | 8 | 4.21 | 2.89 | 141 | −0.17 | 0 |
| Classical disparity filter | 404 | 376 | 4.46E−03 | 12 | 4.39 | 1.86 | 87 | −0.20 | 0 |
| Multilevel disparity filter | 332 | 373 | 6.79E−03 | 12 | 4.44 | 2.25 | 78 | −0.19 | 0 |
| World Air Transportation | 2518 | 16,313 | 5.15E−03 | 9 | 3.62 | 12.96 | 242 | −0.07 | 0.47 |
| Classical disparity filter | 1433 | 4894 | 4.77E−03 | 8 | 3.74 | 6.83 | 120 | −0.15 | 0.16 |
| Multilevel disparity filter | 1470 | 4892 | 4.53E−03 | 8 | 3.86 | 6.66 | 107 | −0.15 | 0.18 |
| Scientific Collaboration | 16,264 | 47594 | 3.60E−04 | 18 | 6.63 | 5.85 | 107 | 0.18 | 0.64 |
| Classical disparity filter | 9264 | 14278 | 3.33E−04 | 26 | 9.68 | 3.08 | 38 | 0.18 | 0.33 |
| Multilevel disparity filter | 9498 | 14217 | 3.15E−04 | 25 | 8.57 | 2.99 | 40 | 0.16 | 0.28 |

*V* is the number of total nodes. *E* is the number of edges. *ω* denotes density. *d* is the diameter. *L* represents the average shortest path. ⟨*k*⟩ is the average node degree. $k_{max}$ represents the max node degree. *r* is the assortativity coefficient. *C* denotes Clustering Coefficient both backbones. For the Madrid Train Bombing network, the Multilevel backbone preserves the diameter and average shortest path, while the Classical backbone preserves the clustering coefficient. The density, average node degree, max node degree, and assortativity coefficient values are similar for both backbones. In the Les Misérables network, the Classical Disparity Filter maintains the density and assortativity coefficient more effectively than the Multilevel backbone. However, the Multilevel backbone is closer to the original network regarding diameter and clustering coefficient. The other properties show similar values for both backbones. Moving on to the Unicode Languages network, the Classical backbone only preserves the density, while the Multilevel backbone only preserves the average node degree of the original network. Both extracted backbones have similar values for all other properties. For the World Air Transportation network, the Classical and Multilevel Disparity Filters exhibit similar values for all the calculated properties. In the Scientific Collaboration network, the Classical Disparity Filter preserves the density, average node degree, and assortativity coefficient, whereas the Multilevel version only preserves the average shortest path of the original network. The diameter, max node degree, and clustering coefficient have comparable values for both Classical and Multilevel backbones. Overall, it can be concluded that the Multilevel Disparity Filter enhances the preservation of global properties compared to the Classical version.

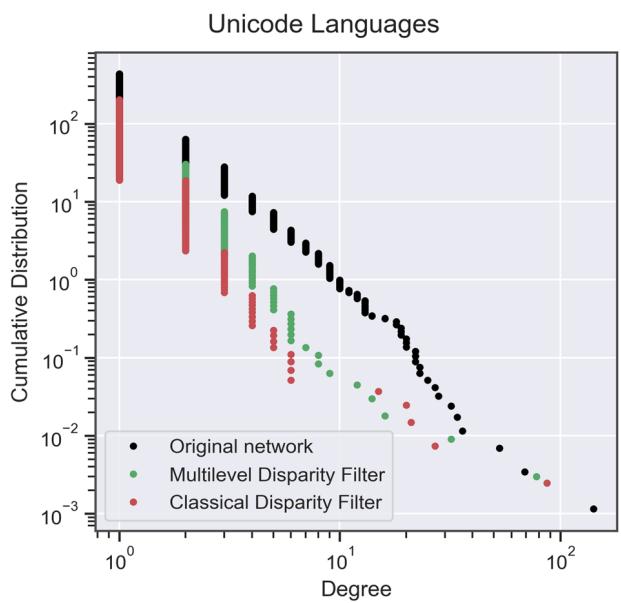

**Fig. 13** Cumulative degree distribution for the Unicode Language network and its extracted Multilevel and Classical Disparity Filter

*Comparing the distributions*

*Cumulative degree distribution* Using a fixed edge percentage of 30%, Fig. 13 illustrates the degree distribution of the Unicode Languages network and its extracted backbones. It is worth noting that the Multilevel disparity filter backbone's cumulative degree distribution is slightly closer to the initial network distribution than the Classical Disparity Filter distribution. The Kolmogorov-Smirnov (KS) statistic between the original backbone distribution and the backbone distribution, as presented in Table 8, highlights this result. The KS statistic is minimal (0.063) for the Multilevel Disparity Filter, indicating a close similarity between its degree distribution and that of the original network. However, the Classical Disparity Filter's degree distribution differs significantly from the original network's degree distribution. Some other networks display similar behavior when comparing the distributions of the Classical and Multilevel Disparity filters. Specifically, the Zachary's Karate Club and World Air Transportation networks exhibit small KS values calculated between the Multilevel Disparity Filter degree distribution and the original network degree distribution. In contrast, the degree distributions of the Classical Disparity Filter for the other four networks are closer to those of the original networks. textitCumulative weight distribution The Multilevel Disparity Filter retains practically all scales of weights, as observed in Fig. 14. In contrast, the Classical Disparity Filter removes several regions of tiny weighted edges. This finding is corroborated by the KS statistic, which indicates that the Multilevel Disparity Filter weight distribution more closely follows the original network weight distribution than the Classical Disparity Filter weight distribution. Similar results have been observed in other networks under investigation (refer to the Supplemental Materials). The original networks' cumulative weight distribution and backbones demonstrate the same behavior as the Unicode Languages network. Except for the Madrid Train Bombing and Les Misérables networks, the KS statistic of the Classical Disparity Filter is somewhat lower than the KS statistic of the Multilevel Disparity Filter, as shown in Table 8. Consequently, the cumulative weight distribution of the Classical Disparity Filter is slightly closer to the original network compared to the Multilevel version.

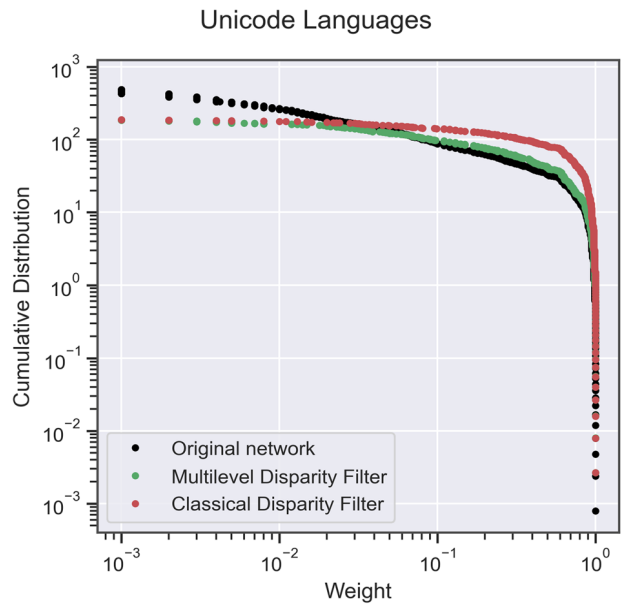

**Fig. 14** Cumulative weight distribution for the unicode language network and its extracted multilevel and classical disparity filter

**Table 8** KS statistics between the backbones degree distribution and the original network degree distribution, as well as between the backbones weight distribution and the original network weight distribution

| Network | Backbone | KS (degree distribution) | KS (weight distribution) |
| --- | --- | --- | --- |
| Zachary's karate club | Classical disparity filter | 0.447 | 0.513 |
| | Multilevel disparity filter | 0.409 | 0.470 |
| Wind surfers | Classical disparity filter | 0.767 | 0.642 |
| | Multilevel disparity filter | 0.806 | 0.464 |
| Madrid train bombing | Classical disparity filter | 0.370 | 0.278 |
| | Multilevel disparity filter | 0.414 | 0.283 |
| Les Misérables | Classical disparity filter | 0.332 | 0.447 |
| | Multilevel disparity filter | 0.384 | 0.480 |
| Unicode languages | Classical disparity filter | 0.149 | 0.495 |
| | Multilevel disparity filter | 0.063 | 0.361 |
| World air transportation | Classical disparity filter | 0.185 | 0.379 |
| | Multilevel disparity filter | 0.157 | 0.378 |
| Scientific collaboration | Classical disparity filter | 0.268 | 0.398 |
| | Multilevel disparity filter | 0.304 | 0.369 |

### Comparing the distances

Turning our attention to various distance metrics, Fig. 15 illustrates the evolution of Portrait Divergence Distance (Left), Laplacian Spectrum Distance (Middle), and Network Laplacian Spectral Distance (Right) between the Unicode Languages network and its extracted backbones in the function of preserved edges. Examining the left panel, we observe that the Portrait Divergence Distance between the original network and the Multilevel Disparity Filter is consistently smaller for all link fractions below 90%. This indicates that the Multilevel backbone more closely resembles the visual properties of the original network compared to the Classical backbone. However, for link fractions between 20% and 60%, the Multilevel Disparity Filter exhibits closer proximity to the original network than the Classical Disparity Filter in terms of the Laplacian Spectrum Distance, as seen in the middle panel. Analyzing the right panel, we can conclude that for fractions of preserved edges below 25% and above 55%, the Multilevel Disparity Filter is more similar to the original network's structure than the Classical Disparity Filter, based on the NetLSD metric. Overall, considering all distance measurements together, the Multilevel Framework utilizing the Disparity Filter algorithm preserves more original graph features compared to the Classical method.

Regarding the other networks, several differences are observed. For the Wind Surfers network, the Laplacian Spectrum Distance between the original network and the Classical Disparity Filter is less than the distance between the original network and the Multilevel Disparity Filter for fractions of preserved edges between 20% and 60%. However, except for fractions between 70% and 80% for Laplacian Spectrum Distance, the extracted Multilevel Disparity Filter exhibits closer proximity to the Madrid Train Bombing network. For all fractions higher than 20%, the Multilevel Disparity Filter closely resembles the original network, according to the Network Laplacian Spectrum Distance. Regarding the Network Laplacian Spectrum Distance calculated for the Les Misérables network, the Classical Disparity Filter is the closest to the original network. However, the Multilevel Disparity Filter appears closer to the original network in terms of the other two distances. It is noteworthy that for the World Air Transportation network, the Classical Disparity Filter tends to be closer to the original network after maintaining 60% of links, in terms of the Network Laplacian Spectrum Distance.

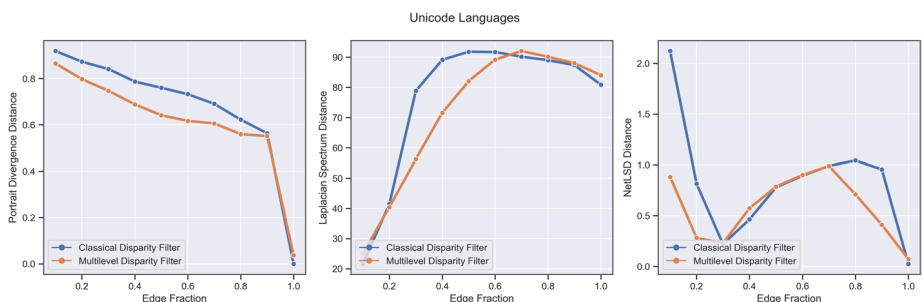

**Fig. 15** Distance measurements according to the fraction of preserved links between Unicode Languages network and Multilevel Disparity Filter (curve in orange), and between Unicode Languages network and Classical Disparity Filter (curve in blue). Panels represent respectively Portrait Divergence Distance (Left), Laplacian Spectrum Distance (Middle), and Network Laplacian Spectral Distance (Right)

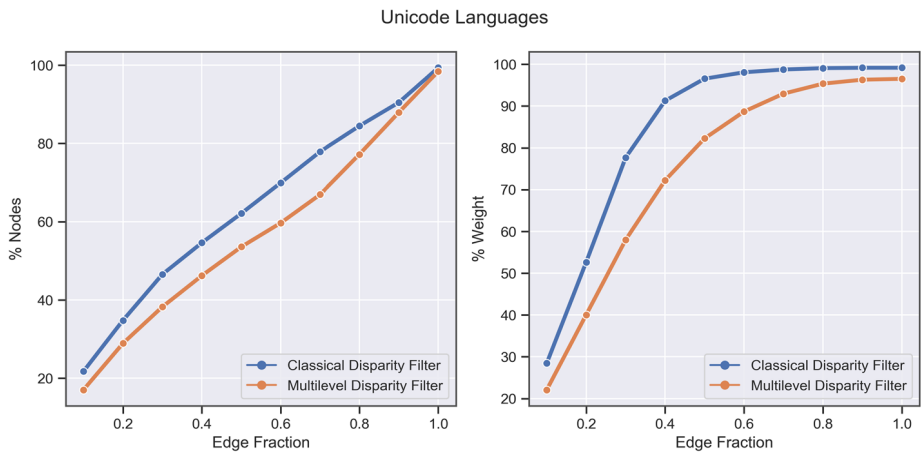

**Fig. 16** Fraction of preserved nodes and weights as a function of the fraction of edges maintained by the multilevel and classical Disparity Filter of Unicode Language network

**Table 9** Modularity of the real-world networks under study and their classical disparity threshold and multilevel disparity threshold

| Network | Modularity | | |
| --- | --- | --- | --- |
| | Original network | Classical disparity filter | Multilevel disparity filter |
| Zachary's karate club | 0.443 | 0.460 | 0.486 |
| Wind surfers | 0.371 | 0.434 | 0.372 |
| Madrid train bombing | 0.435 | 0.392 | 0.402 |
| Les Misérables | 0.565 | 0.550 | 0.519 |
| Unicode languages | 0.771 | 0.802 | 0.743 |
| World air transportation | 0.630 | 0.404 | 0.450 |
| Scientific collaboration | 0.873 | 0.916 | 0.889 |

***Comparing the mesoscopic properties***

textitQuantitative analysis In Fig. 16, we present statistics regarding the relative sizes of retained backbones for the Unicode Languages network when filtered by the Multilevel Disparity Filter and the Classical Disparity Filter in terms of fractions of total weight and nodes. A noteworthy difference is observed as the Multilevel Disparity Filter maintains fewer nodes and lower weights than the Classical Disparity Filter. This behavior can be attributed to the Multilevel Disparity Filter's interest in both large and small weights. Similar patterns are observed in other networks concerning the preserved weight fraction. Moreover, for the Madrid Train Bombing, Les Misérables, and Scientific Collaboration networks, the Multilevel Disparity Filter preserves a higher fraction of nodes compared to the Classical version.

textitModularity Table 9 illustrates the modularity of the investigated real-world networks and their identified backbones through the application of the Disparity Filter. The Classical Disparity Filter, in most instances, increases the modularity of the original network's value, indicating an enhancement of the network's community structure. This is particularly evident in networks such as Wind Surfers, where the modularity increases from 0.371 to 0.434, and in the Scientific Collaboration network, which rises from 0.873 to 0.916. Such increases suggest that the Classical Disparity Filter may reinforce the definition of community boundaries within these networks. Conversely, the Multilevel Disparity Filter tends to result in modularity values closer to the original network's modularity. For instance, in the Wind Surfers network, the modularity is reduced from 0.371 in the original to 0.372 with the Multilevel Disparity Filter, compared to a decrease to 0.434 with the Classical Disparity Filter. In the Madrid Train Bombing network, the original modularity is 0.435, and the Multilevel Disparity Filter produces a modularity of 0.402, closer than the Classical's 0.392.

In conclusion, while the Classical Disparity Filter often enhances the modularity, potentially sharpening the community structure, the Multilevel Disparity Filter seems to preserve the original network's modularity better. This preservation is crucial when we aim to maintain the existing network structure while still identifying significant community divisions. The Multilevel Disparity Filter, therefore, may be preferable in scenarios where a less invasive modification to the network's original structure is desired.

textitParticipation coefficient From Fig. 17, it is observable that the original network exhibits a specific profile of the Participation Coefficient, which serves as a baseline for comparison. The Classical Disparity Filter shows a skewed distribution towards higher Participation Coefficient values, suggesting that it may be more likely to retain or even accentuate a node's connections across multiple communities than the original network. In contrast, the Multilevel Disparity Filter displays a distribution that closely tracks the original network's profile, indicating that it preserves the original network's structure of node connections among communities. The alignment of the Multilevel Disparity Filter with the original network is particularly evident in the central peak of the distribution, where it closely matches the original network's density and range of Participation Coefficient values.

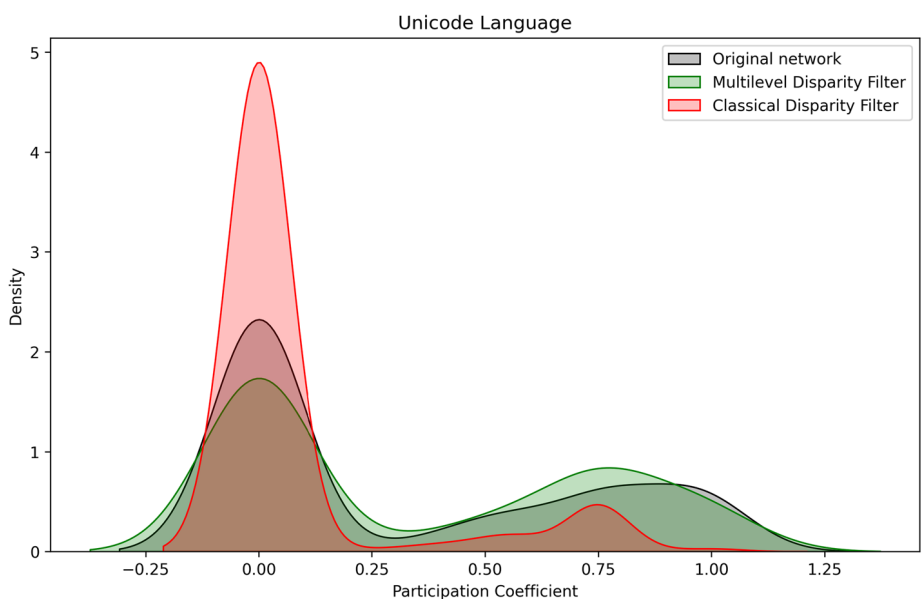

**Fig. 17** Participation coefficient for the unicode language network and its extracted multilevel and classical disparity filter

Based on this comparative analysis, we can deduce that the Multilevel Disparity Filter is more effective in maintaining the Participation Coefficient distribution that represents the original network. This suggests that the Multilevel Disparity Filter would be preferable for analyses requiring the preservation of the original network's community interaction structure. It offers a more faithful replication of the original network's modular connectivity, an essential characteristic for understanding the complex interplay of nodes within and between communities in the network.

textitInter-community connectivity In this part, we compare the impact of the Classical Disparity Filter and the Multilevel Disparity Filter on the Inter-Community Connectivity of a set of real-world networks (refer to Table 10). The goal is to evaluate which filtering method best preserves the inter-community connections as evidenced in the original networks.

The original Inter-Community Connectivity stands at 51.00 for Zachary's Karate Club network. Post-filtering, the Classical Disparity Filter reduces this value to 19.00, while the Multilevel Disparity Filter further reduces it to 18.00. In the case of the Wind Surfers network, the original connectivity is significantly reduced from 118.00 to 11.00 by the Classical Disparity Filter, whereas the Multilevel Disparity Filter lessens the reduction, maintaining a higher connectivity value of 59.00. In the Madrid Train Bombing network, the original Inter-Community Connectivity value of 55.00 is decreased to 14.00 by the Classical Disparity Filter and to a slightly higher value of 17.00 by the Multilevel Disparity Filter. For Les Misérables, the original value of 154.00 drops to 69.00 with the Classical Disparity Filter and is better preserved at 99.00 with the Multilevel Disparity Filter. Starting with connectivity of 37.53, the Unicode Languages network sees a reduction to 21.45 with the Classical Disparity Filter and, to a lesser extent, with the Multilevel Disparity Filter, which yields a value of 30.30. In contrast, the Classical Disparity Filter maintains a higher connectivity value for the World Air Transportation network than the Multiscale Disparity Filter, significantly reducing the original connectivity. Finally, in the case of the Scientific Collaboration network, which has a high original connectivity value of 3001.50, both methods significantly reduce this value. However, the Multilevel

**Table 10** Inter-Community Connectivity of the real-world networks under study and their Classical Disparity Filter and Multilevel Disparity Filter

| Network | Inter-community connectivity | | |
|---|---|---|---|
| | Original network | Classical disparity filter | Multilevel disparity filter |
| Zachary's karate club | 51.00 | 19.00 | 18.00 |
| Wind surfers | 118.00 | 11.00 | 59.00 |
| Madrid train bombing | 55.00 | 14.00 | 17.00 |
| Les Misérables | 154.00 | 69.00 | 99.00 |
| Unicode languages | 37.53 | 21.45 | 30.30 |
| World air transportation | 1,041,054 | 878,974 | 677,358 |
| Scientific collaboration | 3001.50 | 1093.73 | 1497.25 |

Disparity Filter retains higher connectivity (1497.25) than the Classical Disparity Filter (1093.73).

Through this analysis, it emerges that while both the Classical Disparity Filter and the Multilevel Disparity Filter decrease the Inter-Community Connectivity relative to the original networks, the Multilevel Disparity Filter consistently maintains a closer connectivity value to the original network across most of the networks analyzed. These observations are integral to network analysis, especially when the objective is to discern the robustness of the network structure and the efficiency of information or influence spread across communities. Therefore, the Multilevel Disparity Filter can be recommended for preserving the essence of the original network's inter-community interactions.

textitIntra-community connectivity Table 11 provides the Intra-Community Connectivity values of the networks under study and their extracted backbones using Classical Disparity Filter and Multilevel Disparity Filter. Zachary's Karate Club experiences a decrease in Intra-Community Connectivity from the original 180.00 to 79.00 when both the Classical Disparity Filter and Multilevel Disparity Filter are applied, indicating no variance between the two methods in terms of preserving internal community connections. For the Wind Surfers network, the original Intra-Community Connectivity of 1088.00 is reduced to 838.00 by the Classical Disparity Filter, while the Multilevel Disparity Filter results in a slightly lower connectivity of 746.00. Here, the Classical Disparity Filter maintains a connectivity value closer to the original. The Madrid Train Bombing network's connectivity decreases from 227.00 to 98.00 using the Classical Disparity Filter, which is marginally better preserved than the Multilevel Disparity Filter's reduction to 94.00. Les Misérables shows a similar trend; the original connectivity of 666.00 is better preserved by the Classical Disparity Filter, which decreases it to 427.00, compared to the Multilevel Disparity Filter's reduction to 400.00. A pronounced difference is observed in the Unicode Languages network, where the original connectivity of 265.15 is decreased less by the Classical Disparity Filter to 213.63, as opposed to the Multilevel Disparity Filter's reduction to 145.20, indicating a significant advantage for the Classical method in maintaining connectivity similar to the original network. The World Air Transportation network makes an exception in terms of intra-community connectivity. The original connectivity decreases using both methods. However, the Multilevel Disparity Filter preserves a higher connectivity value than the Classical

**Table 11** Intra-community connectivity of the real-world networks under study and their classical disparity filter and multilevel disparity filter

| Network | Intra-community connectivity | | |
|---|---|---|---|
| | Original network | Classical disparity filter | Multilevel disparity filter |
| Zachary's karate club | 180.00 | 79.00 | 79.00 |
| Wind surfers | 1088.00 | 838.00 | 746.00 |
| Madrid train bombing | 227.00 | 98.00 | 94.00 |
| Les Misérables | 666.00 | 427.00 | 400.00 |
| Unicode languages | 265.15 | 213.63 | 145.20 |
| World air transportation | 3,475,074 | 2,259,421 | 2,396,006 |
| Scientific collaboration | 24208.00 | 15598.06 | 14639.58 |

Disparity Filter. Finally, the Scientific Collaboration network, which starts with an intra-community connectivity of 24208.00, retains 15598.06 with the Classical Disparity Filter, whereas the Multilevel Disparity Filter reduces it further to 14639.58, suggesting that the Classical Disparity Filter more closely preserves the network's original connectivity.

Our findings suggest that while both disparity filters reduce Intra-Community Connectivity from the original network values, the Classical Disparity Filter generally preserves this connectivity better than the Multilevel Disparity Filter. This observation is consistent across most networks studied, with the Classical Disparity Filter demonstrating a closer approximation to the original network's intra-community structure. This information is particularly valuable when choosing a backbone extraction method that minimizes the loss of community cohesion within the network.

## Discussion

In this study, using a dataset of seven networks from different domains of various sizes, we use several backbone quality measures to compare our approach with the Classical technique. The comparison's objective is to assess the Backbone extraction Framework based on the Component structure. In the first part, we use a global algorithm to extract the Global Threshold backbone. Results show that the Multilevel Global Threshold is the most effective in weight and degree distributions. The Multilevel Global Threshold maintains all scales of weight. However, the Classical Global Threshold prioritizes edges with significant weights. Therefore, the proposed method can preserve the network structure. In contrast to the Classical method, it does not consider the preservation of the network structure. The backbone may lose one or more components. Furthermore, the Multilevel Global Threshold exhibits good performance in terms of distance. We find small values of distances between the Multilevel Global Threshold and the original network compared with the Classical Global Threshold. It can be concluded that the proposed method is closer to the original network than the Classical method. However, comparing the mesoscopic properties, our findings indicate that while the Classical Global Threshold tends to preserve a higher fraction of total weight, suggesting its preference for stronger links, the Multilevel Global Threshold is more conservative in its approach, often maintaining closer modularity to the original network. The Multilevel Threshold also illustrates this better preserves intercommunity connectivity, particularly in networks with robust initial community structures. The choice of method thus hinges on the specific attributes of the network and the desired outcome of the analysis, whether it is the strength of connectivity or the fidelity to the original community structure that is of primary concern.

The differences in network properties such as diameter, average shortest path, and clustering coefficient between the original networks and their backbones, as extracted using the Multilevel Global Threshold method, can be attributed to both the selected threshold and the community detection method used. The threshold determines which edges are preserved, which directly impacts these properties. Meanwhile, the community detection method influences the initial segmentation of the network, thus affecting the overall structure of the backbone. Our analysis indicates that these variations arise mainly from the interaction between these two factors. The community detection method establishes the initial dense regions, while the threshold defines the edges to be retained in the final backbone. This interaction is crucial for preserving the backbone's integrity and accurately reflecting the network's complexity.

In the second part, we use a local algorithm to extract the backbone, the Disparity Filter. The results indicate that the Multilevel Disparity Filter is the most efficient in terms of weight and degree distributions. The Multilevel Disparity Filter maintains all weight scales. The Classical Disparity Filter focuses on edges with large weights. Furthermore, the network structure may be preserved using the suggested method. Unlike the Classical method, the preservation of the network structure is not considered. The backbone can lose one or more of the network components. Moreover, the Multilevel Disparity Filter performs well in terms of distance. Compared to the Classical Disparity Filter, we obtain low values of distances between the Multilevel Disparity Filter and the original network. In summary, compared to the Classical technique, the suggested method more closely preserves some properties and structure of the network. In our comparative analysis of network structures, the Classical Disparity Filter and Multilevel Disparity Filter exhibit distinct tendencies in terms of preserving intra-community and inter-community connectivity. The Classical Disparity Filter generally maintains a greater degree of the original network's connectivity within communities, suggesting its potential to preserve stronger intra-community ties. On the other hand, the Multilevel Disparity Filter tends to align more closely with the original network's community structure, particularly in preserving the diversity of node connections across different communities. This is reflected in its more conservative approach to reducing inter-community connectivity, maintaining the integral network structure. Both filters, therefore, present unique advantages: The classical Disparity Filter could be preferable for studies emphasizing internal community strength, while the Multilevel Disparity Filter may be better suited for analyses that require a broader view of the network's overall community framework.

The observed variations in distance conservation across different networks and link fractions can be attributed to these networks' inherent structural characteristics and weight distributions. Our empirical results demonstrate that these variations are predominantly influenced by the interplay between the chosen backbone extraction method and the network's structural properties. For example, in networks like the Unicode Languages network, both methods preserved distance metrics relatively well at higher link fractions. However, in more complex and heterogeneous networks such as the World Air Transportation network, the Disparity Filter method showed better performance in maintaining distance metrics.

In this paper, we are analyzing our Multilevel Backbone Extraction Framework based on the Component structure in the context of global and local backbone extraction methods (Global Threshold and Disparity Filter) to reveal distinct aspects of network structure that each method can preserve. For the Global method, it appears to be effective in preserving global properties such as network diameter and density, particularly in larger networks. This method also maintains the degree and weight distributions across the network, ensuring minimal distance alteration between the backbone and the original network, which is crucial for maintaining the integrity of the network's structure. However, for the Local method, represented by the Disparity Filter, shows efficacy in preserving the weight distribution and minimizing the distance from the original network structure. Regarding mesoscopic properties, it effectively maintains modularity, participation coefficient, and inter-community connectivity, especially in larger networks, with a minimal total number of nodes and weights.

When considering mesoscopic properties, both methods show strengths in different areas. The Global method exhibits robustness in maintaining modularity, participation coefficient, and inter-community connectivity, particularly for the largest network analyzed, along with minimal alteration to the total number of nodes and weights. This suggests that the Global Threshold method might be more suitable for preserving community structures for larger-scale networks. Conversely, the local method improves mesoscopic properties for larger networks, particularly in modularity, participation coefficient, and inter-community connectivity. This indicates the method's ability to maintain the community-based structure and the interaction between these communities within the network.

## Conclusion

The architecture and dynamics of complex systems must be characterized through analysis. However, this procedure is limited by the complexity of networks. It is crucial to eliminate redundant data from the network while maintaining nodes and edges that preserve pertinent information (Yassin et al. 2022a). As a result, scientists are interested in backbone extraction or filter-based methods to solve this problem. This paper investigates a new approach to backbone extraction based on component structure.

This work reveals a comparative study between the Multilevel and Classical backbone extracted from real-world networks. The basic global properties, the weight and degree distributions, the mesoscopic properties, and the distances are used to evaluate the proposed method. Results show that the proposed backbone extraction framework is the most effective method to extract the backbone while preserving the mesoscopic representation of the network compared to Classical methods. This finding is confirmed regarding distributions, mesoscopic properties, and distance measures. It's not the case for the basic global properties. Results show that these properties are enhanced in the Classical version.

In conclusion, the suggested framework successfully exploits the heterogeneity of real-world networks, improving the benefits of Classical methods. Future research will examine its association with more advanced backbone extraction methods. We also intend to examine the effects of different component structure detection techniques.

## Supplementary Information

The online version contains supplementary material available at https://doi.org/10.1007/s41109-024-00645-z.

> Supplementary Material 1.

**Acknowledgements**
This research is supported by the project "Plateforme logicielle d'intégration de stratégies d'immunisation contre la pandémie COVID-19" funded by the grant of the Hassan II Academy of Sciences and Technology of Morocco.

**Author contributions**
SH performed the experiments and wrote the first draft. All authors conceived the study and discussed the results and edited the manuscript.

**Funding**
Not applicable

**Availibility of data and materials**
This manuscript does not report data generation or analysis.

## Declarations

### Ethics approval and consent to participate
Not applicable.

### Consent for publication
Not applicable

### Competing interests
The authors declare that they have no competing interests.

## Publisher's Note